\documentclass[preprint,final,11pt]{elsarticle}%

\usepackage{amsmath}
\usepackage{subfig}
\usepackage{geometry}
\usepackage{amssymb}
\usepackage{amsthm}
\usepackage{amsfonts}
\usepackage{amsmath}  
\usepackage{graphicx}
\usepackage{setspace}
\usepackage{fancyhdr}
\usepackage{xcolor}
\usepackage{lineno}
\usepackage{caption}
\usepackage[numbers]{natbib}
\usepackage{multirow}%
\usepackage{subfig}%
\usepackage[utf8]{inputenc}
\usepackage{amsmath}
\usepackage{amssymb}
\usepackage{amsfonts}
\usepackage{float}
\usepackage{mathtools}
\usepackage{tikz}
\usepackage{pgfplots}
\usepackage{tikzscale}
\usepackage{graphicx}
\pgfplotsset{compat=1.14}

\newcommand{\ten}[1]{\boldsymbol{#1}}

\begin{document}

\begin{frontmatter}
\title{Data-driven, structure-based hyperelastic manifolds: \\ A
macro-micro-macro approach}
\author{V\'{i}ctor Jes\'{u}s Amores}
\author{Jos\'{e} Mar\'{i}a Ben\'{i}tez}
\author{Francisco Javier Mont\'{a}ns \corref{cor1}}
\ead{fco.montans@upm.es}
\cortext[cor1]{Corresponding author}
\address{Escuela T\'{e}cnica Superior de Ingenier\'{\i}a Aeron\'{a}utica y del Espacio\\Universidad Polit\'{e}cnica de Madrid\\
Plaza Cardenal Cisneros, 3, 28040-Madrid, Spain}
\begin{abstract}
In this paper we introduce a novel approach to obtain the stored energy density of rubber-like materials directly from experimental data. The model is structure-based, in which the only assumption is the existence of an isotropic distribution of fibres, chains or networks. Using a single macroscopic test, we obtain the response of the constituents simply by solving a linear system of equations. This response includes all possible interactions, without an assumption on the nature of that behavior, performing an efficient reverse-engineering of the constituents behavior. With that microstructural behavior, we build constitutive manifolds capable of reproducing accurately the behavior of the continuum  under any arbitrary loading condition. To demonstrate the goodness of the proposed non-parametric macro-micro-macro approach, using just one of the macroscopic test curves of the Kawabata et al experiments to compute the fiber (micro) behavior,  we reproduce, to very good accuracy, the rest of the series of biaxial tests from Kawabata on vulcanized rubber 8phr sulfur, showing that the computed micromechanical behavior represents an excellent approximation of the actual behavior. We show similar results for the Treloar material and for the Kawamura et al series of experiments on silicone. With the use of constitutive manifolds, the method has similar efficiency in finite element programs to that of analytical models. Julia language code to reproduce the main computations and some figures of the paper is available.

\end{abstract}
\end{frontmatter}

\section{Introduction}

The behavior of polymeric-like materials, including elastomers and soft biological tissues, is characterized by a highly nonlinear elastic behavior capable of developing very large strains without dissipating a relevant amount energy \cite{mark2007rubberlike,benitez2017mechanical,bathe2006finite,treloar1975physics,bergstrom2015mechanics}. 
As a consequence of this conservative behavior, these materials are modeled as hyperelastic materials by assuming the existence of a potential function, also called strain energy function \cite{benitez2017mechanical,bergstrom2015mechanics}. The mathematical setting of the potential has been traditionally performed inferring an analytical function shape, a shape which accommodates experimental evidence through some material parameters. These material parameters are determined by fitting the model predictions to the experimental data from the material by means of nonlinear optimization techniques, where non-uniqueness of the parameters solution in these procedures may be a relevant issue \cite{latorre2017understanding,ogden2004fitting}. We classify these hyperelastic models in three categories as follows.

\subsection{Macro (or phenomenological) approaches} Examples of the phenomenological approach are the Ogden model, the Gent model, the Yeoh model, etc. The main advantage of the phenomenological (macro) approach is that there is no need to make assumptions on the microstructure or its behavior. However, the accuracy of these models under general deformations depends on the selected analytical shape, on the selected invariants (which govern the uncoupling assumptions of the energy terms) and on the number of material parameters. In general, these models require (or should require) a larger number of experimental data than the structure-based ones, and result in increased complexity for increased accuracy. Hence, data-driven approaches have been proposed to increase the generality in modelling and to incorporate data in a more direct manner \cite{ibanez2017data,ibanez2018hybrid,ibanez2018manifold}. Within the phenomenological framework, and based on the ideas of the Sussman-Bathe model  for isotropic, incompressible materials \cite{sussman2009model}, we have pursued a non-parametric data-driven approach where neither an analytical shape of the energy function nor material parameters are considered. Due to the  data-driven conception and the ability to accurately capture experimental data, we named this new approach as WYPiWYG (What-You-Prescribe is What-You-Get) hyperelasticity.  WYPiWYG hyperelasticity is a procedure that, instead of fitting, \emph{solves} numerically the equilibrium differential equations that govern a given test, obtaining the material behavior for more general cases. Then, a continuous stored energy density for finite element analysis may  be  obtained interpolating the solved discrete values using, for example, splines or B-splines. WYPiWYG hyperelasticity has been successfully applied to anisotropic materials \cite{latorre2013extension,latorre2014you}, compressible materials \cite{crespo2017wypiwyg}, and unconventional materials \cite{crespo2018continuum}. In all these cases, the proposed models have been able to reproduce smooth experimental tests to any desired precision. Similar accuracy has been found for capturing a variety of loading cases \cite{latorre2018continuum} or when WYPiWYG models have been implemented in finite element codes and compared to analytical ones \cite{crespo2017wypiwyg,de2017capturing}. In the case of noisy experiments, stability conditions may be guaranteed when they are needed \cite{latorre2018experimental}.
\subsection{Micro-macro (or classical structure-based, bottom-up) approach}Examples of micro-macro approaches for isotropic materials are the eight-chain model, the extended tube model and Miehe's microsphere model. These models take into account the molecular structure of the material by means of functions and micro-mechanical material parameters which are related to that micro-structure. These functions are based on \textit{assumptions} about the behavior of the constituents (e.g. free-joint chains, transverse constraints, etc). Indeed, the most common procedure is based on statistical mechanics, assuming that the components are freely joined chains composed of $N$ segments of equal length, called Kuhn segments. Each Kuhn segment has a length $l $, so the length of the chain is $L=Nl$. The chain  is \textit{assumed} rigid and coiled but free to rotate, keeping the internal energy unchanged in an entropy-governed setting. If we also assume no intermolecular interaction, the stored energy is obtained by the sum of the energy stored in each chain \cite{treloar1975physics,bergstrom2015mechanics}.

One of the best-known models of this kind is the eight-chain affine model developed by Arruda and Boyce. The reason for the success of this model in engineering practice is that with only two material parameters, easily obtained from a tensile test, the model is capable of reproducing the behavior of isotropic incompressible materials to some accuracy as shown in \cite{arruda1993three} for the Treloar tests \cite{treloar1944stress}. However, when obtaining the parameters from a tensile test, the stresses for the equibiaxial test are understimated. It is frequently assumed that the inability of this model to capture the equibiaxial test with parameters from the tensile test is that the influence of the second invariant is neglected (e.g. \cite{bergstrom2015mechanics}). Miehe et al. \cite{miehe2004micro} argued that this underestimation was due to the unconstrained assumption of the chains. Following similar ideas as in the extended tube model \cite{kaliske1999extended}, in their non-affine microsphere model they constrained the movement of a single chain assuming it embedded in a micro-tube. This constraint and the non-affinity insert three  material parameters additional to those of the eight-chain model ($5$ in total). To determine them,  \textit{three} Treloar tests were needed \cite{miehe2004micro}, therefore loosing the main asset of the Arruda-Boyce model. 

Recently, Khi\^{e}m and Itskov observed that the results of this kind of models depended on the numerical integration scheme  \cite{khiem2016analytical}, so to avoid numerical integration they developed a micro-macro mechanical model based on the network-averaging of the tube model with a closed-form of the Rayleigh non-Gaussian distribution. This model was formulated with four material parameters but, again, the set of \textit{three} experimental tests by Treloar  were needed to determine them.
As they noted, citing Urayama \cite{urayama2006experimentalist}, it is believed that several tests are needed to capture the influence of the two invariants (or, say, chain behavior and constraint effect), see also \cite{bergstrom2015mechanics}.
Similar arguments are given in the review paper from Marckmann and Verron \cite{marckmann2006comparison}, where twenty macro and micro-macro models are compared regarding their suitability for predicting both the Treloar and the Kawabata et al experiments. Remarkably, as it can be deduced from Figs. 1 to 8 and Table IV of their paper, the best models in terms of  accuracy are the Shariff \cite{shariff2000strain} and Ogden \cite{ogden1997non} models (phenomenological), which need two curves for characterization. Interestingly, the microsphere model (Fig. 8 of \cite{marckmann2006comparison}), despite accounting for micromechanical chain behavior, transverse chain constraint and non-affinity of deformations, it did not resulted in better accuracy than Ogden's model  (Fig. 3 of \cite{marckmann2006comparison}), and despite employing all that information about the microstructure, it needs to be characterized with three curves (one more than Ogden's model). Hence, \textit{from the modelling point of view}, no practical gain seems to be obtained when using micro-macro models.

To understand the difference between the micro-macro and  the following macro-micro-macro approach, note that, for example, the number of chain segments $N$ and the chain density $n$, are micro-structural material parameters which could be measured analyzing the microstructure; but in practice they are just best-fitted to macroscopic tests. Then, it remains an open question if these parameters (and thus, also the constitutive assumptions) match the actual microstructure, specially when a phenomenological model (which no not use micromechanical insight) performs at least as well using the same or less tests to characterize the material. 
\subsection{Macro-micro-macro (present structure-based) approach }In this paper a new non-parametric data-driven macro-micro-macro approach is proposed for modeling  isotropic hyperelastic materials. In this approach the macroscopic behavior is not obtained from a chain (microscopic) behavior assumption; this latter is obtained directly from the macroscopic one.  Here, the only assumption made is the existence of a network of chains (or fibers). No conjecture about the origin of the behavior of the chains is made. No analytical form of the chain behavior, nor material parameters, are assumed for that behavior. Instead, the behavior of the chain is reverse-engineered directly from a macroscopic test by solving a linear system of equations. Thereafter, the macroscopic behavior for a general loading case is reproduced by the integration of the stored energy of the microscopic chains in a unit sphere, accounting for their isotropic distribution. Directly from macroscopic data, the model implicitly captures the effects of the  micromechanical interactions, cross-linking  and stress-induced crystallization at large strains without any explicit assumption. Pre-integrated macroscopic manifolds may be easily built from the computed micromechanical behavior, so the implementation in finite element codes is standard, straightforward, and as efficient as that of phenomenological models.

Whereas in the micro-macro approach followed by classical structure-based models, material parameters are used to best-fit the \emph{assumed} behavior, in the present macro-micro-macro approach we \emph{solve} numerically for the actual behavior of the microscopic constituents with the single assumption of their isotropic distribution. As a clear advantage of this approach, \textit{we only need one test}, whereas as mentioned, using the classical micro-macro approach the authors usually report to need several sets of experimental data (see e.g. section $6.4$ of \cite{miehe2004micro} and section 3.1 of \cite{khiem2016analytical}).
Then, we obtain the relevant practical advantage that could be expected if the microstructural behavior is properly determined: \textit{a minimum number of tests is needed to model the material in a way that accurately predicts its behavior under any loading condition}.

\subsection{Biaxial tests for analyzing general loading case predictions.}

For an isotropic, incompressible material, biaxial tests with two independent stretches represent all possible loading states, because the third stretch is determined by incompressibility and the material has no directional preference. Then, to demonstrate the accuracy of the obtained micromechanical behavior in  representing the macroscopic material behavior under general loading conditions,  we apply it below to the biaxial test series by   Kawabata et al. \cite{kawabata1981experimental} and Kawamura et al \cite{Kawamura}; and for completeness also to the classical tests from Treloar \cite{treloar1944stress}. As remarked by Khi\^{e}m and Itskov \cite{khiem2016analytical} and shown by Marckmann and Verron \cite{marckmann2006comparison}, only the models by  Kaliske and Heinrich \cite{kaliske1999extended}, Shariff \cite{shariff2000strain}, Miehe et al. \cite{miehe2004micro} and Khi\^{e}m and Itskov \cite{khiem2016analytical}  are able to reproduce with acceptable accuracy the experiments from Kawabata et al \cite{kawabata1981experimental} and Treloar  \cite{treloar1944stress}; the model of Ogden does it using different material parameters for both materials, but Khi\^{e}m and Itskov and Marckmann and Verron argued that they are the same material, so they should have the same behavior. We emphasize that \textit{all }these models need more than one test to obtain the material parameters. We show below that with our proposal we are able to predict to excellent accuracy the experiments of Kawabata et al \cite{kawabata1981experimental} using only the information of a single experimental curve to calibrate the material. In \ref{appen_Tre_kawa} we also show that we are able to predict the experiments of Treloar using the information of a single Treloar test and that the model calibrated with the Treloar material predicts to good accuracy the experiments of Kawabata et al \cite{kawabata1981experimental}, but we indeed observe that both materials do \textit{not} behave in an identical manner during the tests. To further show the applicability to other materials under general deformations, using also one test curve and the same procedure, we also predict the behavior of two silicone materials from Kawamura et al \cite{Kawamura} in a series of biaxial tests, obtaining similar accuracy as for the Kawabata experiments.

Given the numerical emphasis of the data-driven approach,  Julia language code to reproduce the results for the Kawabata and Treloar experiments  in the paper, as well as the figures, is provided as supplementary material.

\section{Proposed macro-micro-macro approach}

Some hyperelastic models,  among them the Arruda-Boyce one, follow a scheme based on the pre-integration of the strain energy function of the chains expressed in terms of an average stretch, see \ref{Sec chain}. Our first approach to the problem followed this path. However,  according to the results shown in \ref{Sec chain}, it seems difficult to obtain an accurate procedure following a macroscopic fitting of a pre-integrated scheme. For this reason, we pursued a different macro-micro-macro approach, in which instead of interpolating and solving for a pre-integrated expression, we interpolate and solve for the fiber function itself by pushing up that interpolation to the continuum level. Then, discrete values of the microstructural behavior are obtained solving a \textit{linear} system of equations at the continuum level, fully determining the needed micro-mechanical behavior upon employing proper interpolation. Thereafter, with this micro-mechanical behavior, constitutive macroscopic manifolds may be construed to determine the behavior of the continuum under any arbitrary deformation directly at that level, without the need of performing numerical integration at the micro level.

\begin{figure}[h!]
\centering
\includegraphics[scale = 0.6]{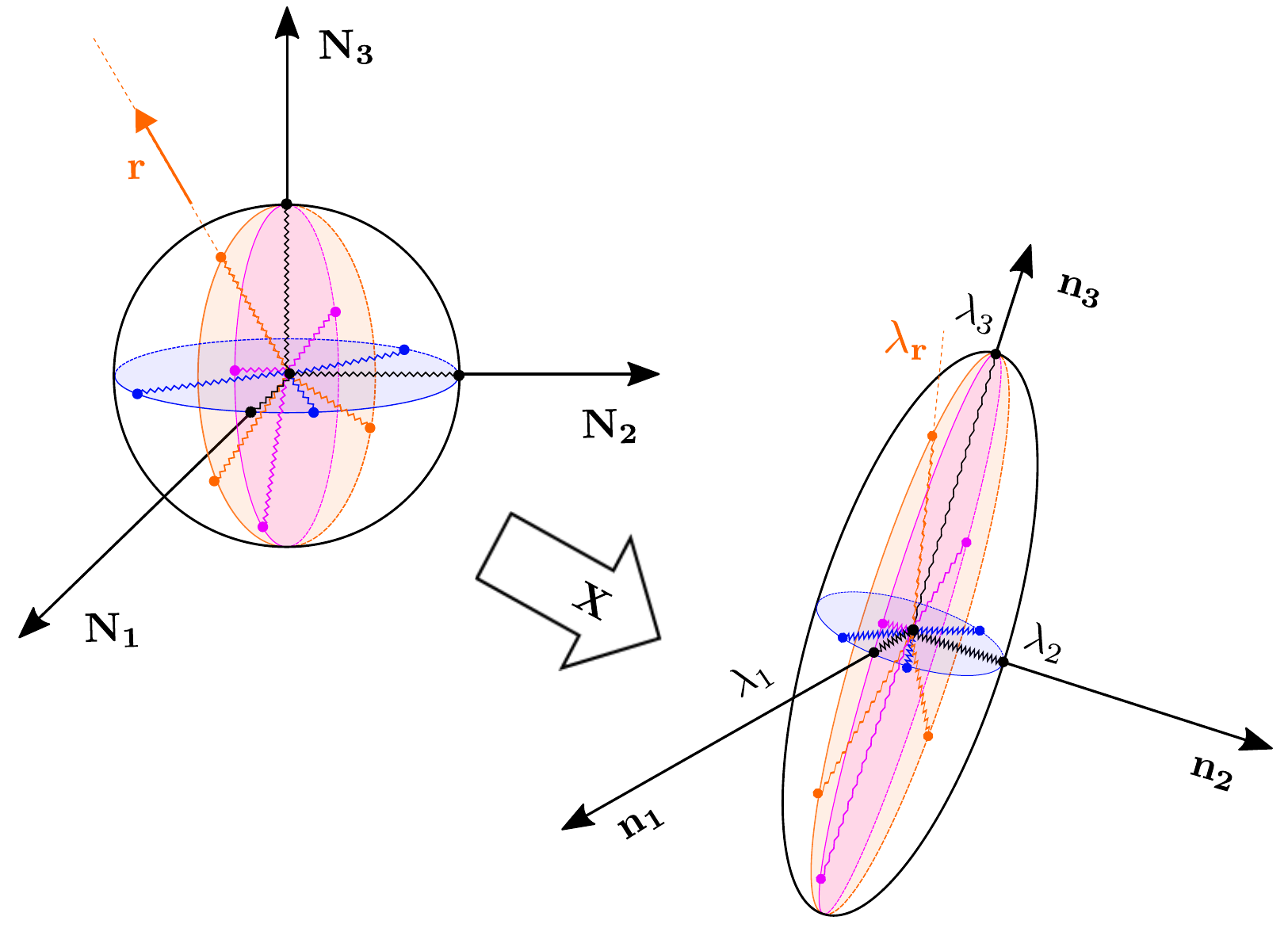}
\caption{Affine deformation of the unit sphere.}
\label{sphere}
\end{figure}

Therefore, our proposal only assumes the existence of chain, fiber-like, or network-like,  elementary
structures (to which we will generically refer to as chains) isotropically distributed in the solid, whose elongation or shortening at the micro-scale level determines
the behaviour of the continuum at the macroscopic level. We assume that the behavior of all chains is the same and that it can be modeled through the potential $\tilde{\Psi}_{ch}(\lambda_{ch})$, where $\lambda_{ch}$ is the stretch of a single chain. Then, given the
stretch of every single chain in the solid, it is
possible to obtain the macroscopic stored energy density function in terms of the principal stretches $\lambda_{1},\lambda_{2}$ and $\lambda_{3}$, i.e. $\Psi(\lambda_{1},\lambda_{2},\lambda_{3})$.  Conversely, if that one-to-one relation holds, it is in general possible to determine the behavior of the representative chain, including all chain interactions, from a single macroscopic test. This is possible because the behavior of the chain depends on a single variable $\lambda_{ch}$, and all chains are equal, and isotropically distributed. Furthermore, since all interactions are included and the functional relation is open (it will be determined below), we can assume an \textit{affine} relation for $\lambda_{ch}$ as $\lambda_{ch}(\lambda_1,\lambda_2,\lambda_3,\theta,\phi)$, where $\theta$ and $\phi$ are the spherical angles giving the orientation of the fiber respect to the principal directions of deformations.  Under the affinity assumption, any sphere in the continuum will be deformed according to the principal stretches, see Fig. \ref{sphere}. From Figure  \ref{sphere}, it is obvious that the stretch in any arbitrary direction of the sphere, $\ten{r}$, is 
\begin{equation}
\lambda_{ch}=\left(  \ten{r\otimes r}\right)  \colon\ten{U=}
\lambda_{1}\sin^{2}\phi\cos^{2}\theta+\lambda_{2}\sin^{2}\phi\sin^{2}
\theta+\lambda_{3}\cos^{2}\phi=\lambda_{1}r_{1}^{2}+\lambda_{2}r_{2}
^{2}+\lambda_{3}r_{3}^{2}\label{rvec}
\end{equation}
where $\ten{U}=\sqrt{\ten{C}}=\sqrt{\ten{X}^{T}\ten{X}}$ is the stretch tensor, $\ten{C}$ is the right Cauchy-Green deformation tensor and $\ten{X}$ is the deformation gradient.
Considering the number of chains in a sphere $n_{ch}$, the
density of chains per volume unit is $\rho=n_{ch}/V$, where $V$ is the volume of the  sphere (e.g. $\tfrac{4}{3}\pi$). The energy density of the continuum is
\begin{equation}
\begin{split}
\Psi(\lambda_{1},\lambda_{2},\lambda_{3})&=
{\displaystyle\int\nolimits_{V}}
\rho\tilde{\Psi}_{ch}(\lambda_{ch})dV=\frac{1}{V_{}}
{\displaystyle\int\nolimits_{V}}
\Psi_{ch}(\lambda_{ch})dV=\frac{1}{S}
{\displaystyle\int\nolimits_{S}}
\Psi_{ch}(\lambda_{ch})dS
\end{split}
\end{equation}where $S$ is the surface of the sphere (e.g. $4\pi$) and  $\Psi_{ch}(\lambda_{ch}):=n_{ch}
\tilde{\Psi}_{ch}(\lambda_{ch})$. If we determine the stored energy density, the nominal stress for an incompressible isotropic material is:

\begin{equation}
P_{i}=\frac{1}{\lambda_{i}}p+\frac{\partial\Psi(\lambda_{1},\lambda
_{2},\lambda_{3})}{\partial\lambda_{i}} \label{Pnom}%
\end{equation}where $p$ is the pressure-like Lagrange multiplier.
Using the chain rule, the derivative of the stored energy is

\begin{equation}
\frac{\partial\Psi(\lambda_{1},\lambda_{2},\lambda_{3})}{\partial\lambda_{i}
}=\frac{1}
{S_{}}
{\displaystyle\int\nolimits_{S}}
\frac{d\Psi_{ch}(\lambda_{ch})}{d\lambda_{ch}}\frac
{\partial\lambda_{ch}}{\partial\lambda_{i}}dS\\
=\frac{1}{S_{}}
{\displaystyle\int\nolimits_{S}}
P_{ch}\left(  \lambda_{ch}\right)  \frac{\partial\lambda_{ch}\left(  \lambda
_{1},\lambda_{2},\lambda_{3},\ten{r}\right)  }{\partial\lambda_{i}}dS\label{eq4}
\end{equation}
where $\ten{r}(\theta,\phi)$ gives the direction of each chain.

Now, our purpose is to determine numerically the function $P_{ch}(\lambda_{ch})$ governing the mechanical behavior directly from data of a macroscopic test, for example a tensile test.
In order to do so, consider the following spline-based interpolation ---see also Section \ref{1Dspline} below
\begin{equation}
P_{ch}\left(  \lambda_{ch}\right)  =
{\displaystyle\sum\limits_{i=1}^{nv}}
N_{i}\left(  \lambda_{ch}\right)  \hat{P}_{ch_{i}}=
\begin{bmatrix}
N_{1}\left(  \lambda_{ch}\right)  & \cdots & N_{nv}\left(  \lambda
_{ch}\right)
\end{bmatrix}
\begin{bmatrix}
\hat{P}_{ch_{1}}\\
\vdots\\
\hat{P}_{ch_{nv}}
\end{bmatrix}
\label{eqPch}
\end{equation}
where $nv$ is the number of B-spline vertices which have prescribed abscissae $\hat{\lambda}_{ch_{1}%
},...,\hat{\lambda}_{ch_{nv}}$, which limits are computed from the available range in the tests and Eq. (\ref{rvec}), and corresponding ordinate  values $\hat{P}_{ch_{1}%
},...,\hat{P}_{ch_{nv}}$, to be computed in this procedure. The functions $N_{i}\left(  \lambda_{ch}\right) $ are the interpolation functions, e.g. the B-splines in the present case; see details in \cite{latorre2018experimental,amores2019ADES}; summarized also in Sec. \ref{1Dspline} below. If we introduce the expression for $P_{ch}\left(  \lambda_{ch}\right)  $ given in Eq. (\ref{eqPch}) into Eq. (\ref{eq4}) we obtain

\begin{equation}
\frac{\partial\Psi(\lambda_{1},\lambda_{2},\lambda_{3})}{\partial\lambda_{i}
}=\frac{1}{S_{}}
{\displaystyle\int\nolimits_{S}}
\left({\displaystyle\sum\limits_{m=1}^{nv}}
N_{m}\left(  \lambda_{ch}\left(  \lambda_{1},\lambda_{2},\lambda
_{3},\ten{r}\right)  \right)  \hat{P}_{ch_{m}}\right)  \frac{\partial
\lambda_{ch}\left(  \lambda_{1},\lambda_{2},\lambda_{3},\ten{r}\right)
}{\partial\lambda_{i}}dS
\end{equation}
where we have left explicit the dependencies for the reader's convenience.

The sphere integral can be computed numerically using a Gaussian quadrature.
The one used here, proposed by Bazant and Oh \cite{bavzant1986efficient}, has $nq$ number of quadrature points (in the examples below we use  the case $nq=21$, the same one used for example by Miehe et al. in their non-affine model \cite{miehe2004micro}):

\begin{equation}
\frac{1}{S_{}}%
{\displaystyle\int\nolimits_{S}}
f(\ten{r})dS\simeq
{\displaystyle\sum\limits_{j=1}^{nq}}
f(\ten{r}_{j})w_{j}
\end{equation}where $w_j$ are the integration weights. Then
\begin{equation}
\frac{\partial\Psi(\lambda_{1},\lambda_{2},\lambda_{3})}{\partial\lambda_{i}}=
{\displaystyle\sum\limits_{j=1}^{nq}}
w_{j} \frac
{\partial\lambda_{ch}\left(  \lambda_{1},\lambda_{2},\lambda_{3}
,\ten{r}_{j}\right)  }{\partial\lambda_{i}}
{\displaystyle\sum\limits_{m=1}^{nv}}
N_{m}\left(  \lambda_{ch}\left(  \lambda_{1},\lambda_{2},\lambda
_{3},\ten{r}_{j}\right)  \right)  \hat{P}_{ch_{m}}  
\end{equation}
Write for simplicity $\lambda_{ch}\left(  \lambda_{1},\lambda_{2},\lambda
_{3},\ten{r}_{j}\right)  =:\lambda_{chj}$ and note that by Eq. (\ref{rvec}) we have  the relation ${\partial\lambda
_{ch}\left(  \lambda_{1},\lambda_{2},\lambda_{3},\ten{r}_{j}\right)
}/{\partial\lambda_{i}}=r^2_{ji}$, which is the square of the $i-th$ component of the vector
$\ten{r}_{j}$ (at integration point $j$ of the quadrature). Since the vector of B-spline vertices $\hat{P}_{ch_{k}}$ is constant, it can be factored-out from the sum. Then
we can write the sum in $m=1,...,nv$  in matrix notation as
\begin{equation}
\frac{\partial\Psi(\lambda_{1},\lambda_{2},\lambda_{3})}{\partial\lambda_{i}}=
\begin{bmatrix}
{\displaystyle\sum\limits_{j=1}^{nq}}
r^2_{ji}w_{j}  N_{1}\left(  \lambda_{chj}\right)   & \cdots &
{\displaystyle\sum\limits_{j=1}^{nq}}
r^2_{ji}w_{j}  N_{nv}\left(  \lambda_{chj}\right)
\end{bmatrix}
\begin{bmatrix}
\hat{P}_{ch_{1}}\\
\vdots\\
\hat{P}_{ch_{nv}}
\end{bmatrix}
\label{dPsidli}\end{equation} which can be written in compact form as

\begin{equation}
\frac{\partial\Psi(\lambda_{1},\lambda_{2},\lambda_{3})}{\partial\lambda_{i}}=
{\displaystyle\sum\limits_{j=1}^{nq}}r^2_{ji}
\begin{bmatrix}
\operatorname{row_N}\left(  \lambda_{chj}\right)
\end{bmatrix}
\begin{bmatrix}
\hat{P}_{ch_{1}}\\
\vdots\\
\hat{P}_{ch_{nv}}
\label{derivativesolution}
\end{bmatrix}
\end{equation}where $\begin{bmatrix}
\operatorname{row_N}\left(  \lambda_{chj}\right)
\end{bmatrix}
=w_{j}  \begin{bmatrix}
N_{1}\left(  \lambda_{chj}\right)  & \cdots & N_{nv}\left(  \lambda
_{chj}\right)
\end{bmatrix}$.  Now, as an  example (a similar procedure is obtained for any other test), using Eq. (\ref{Pnom}) for the case of a biaxial test in the $1-2$ plane $\left(
P_{1},P_2,P_{3}=0\right)  $:
\begin{equation}\label{eq21}
\begin{split}
P_{1}=\frac{\partial\Psi(\lambda_{1},\lambda_{2},\lambda_{3})}{\partial\lambda_{1}}-\frac{\lambda_{3}}{\lambda_{1}
} \frac{\partial\Psi(\lambda_{1},\lambda_{2},\lambda_{3})}{\partial\lambda_{3}}  ={\displaystyle\sum\limits_{j=1}^{nq}}
\left(  r^2_{j1}-r^2_{j3}\frac{\lambda_{3}}{\lambda_{1}}\right) 
\begin{bmatrix}
\operatorname{row_N}\left(  \lambda_{chj}\right)
\end{bmatrix}
\begin{bmatrix}
\hat{P}_{ch_{1}}\\
\vdots\\
\hat{P}_{ch_{nv}}
\end{bmatrix}
\end{split}
\end{equation}
This equation is established for as many points as desired from the experimental data. In practice, from experimental data points, a spline function $P_{1}\left(
\lambda_{1}\right)  $ can be built, and as many points as desired may be obtained by sampling the spline function. Consider a uniaxial test or a test with $\lambda_{2}$ fixed. Taking $nl$ points with uniaxial stretches $\hat{\lambda}_{u1},\cdots,\hat{\lambda}_{unl}$ (note that we need  $nl>nv$), and corresponding nominal stresses $P_1(\hat\lambda_i)$, the previous equation takes the form

\begin{equation}
\begin{bmatrix}
P_{1}(  \lambda_{1}=\hat{\lambda}_{u1}) \\
\vdots\\
P_{1}(  \lambda_{1}\text{=}\hat{\lambda}_{unl})
\end{bmatrix}
\simeq
\begin{bmatrix}
\tilde P_{1}(  \lambda_{1}=\hat{\lambda}_{u1}) \\
\vdots\\
\tilde P_{1}(  \lambda_{1}\text{=}\hat{\lambda}_{unl})
\end{bmatrix}
\equiv
\begin{bmatrix}
\bar N_{1,1} & \cdots & \bar N_{1,nv}\\
\vdots & \cdots & \vdots\\
\bar N_{nl,1} & \cdots & \bar N_{nl,nv}
\end{bmatrix}
\begin{bmatrix}
\hat{P}_{ch_{1}}\\
\vdots\\
\hat{P}_{ch_{nv}}
\end{bmatrix}
\end{equation}where $P_1(\hat\lambda_i)$ are the values from the experimental curve for stretches $\hat\lambda_u$ and $\tilde P_{1}(  \lambda_{1}=\hat{\lambda}_{ui})$ are the corresponding model continuum values obtained from the computed B-spline representing the chain behavior.
 The coefficients $\bar N_{i,j} $ of the matrix are
 \begin{equation}
 \bar N_{i,j}= {\displaystyle\sum\limits_{q=1}^{nq}}
\left(  r^2_{q1}-r^2_{q3}{\frac{\lambda_{3}(\hat\lambda_{ui})}{\hat\lambda_{ui}}}\right)  w_{q}  N_{j}\left(  \lambda_{chq}(\hat\lambda_{ui})\right)
 \end{equation}
 
The previous system of equations is rectangular (usually $nl\gg nv$), and it can be written in the
following compact form:

\begin{equation}
\ten{\tilde P}_{1}(  \ten{\hat{\lambda}}_{u})  =\ten{\bar N}
(  \ten{\hat{\lambda}}_{u})\ten{\hat{P}}_{ch}
\end{equation}
where $\ten{\hat{\lambda}}_{u}$ is the array of uniaxial stretch values, $\ten{\bar N}(  \ten{\hat{\lambda}}_{u})$ is the matrix of coefficients $N_{i,j}$, which depend on the uniaxial stretch values $\hat\lambda_{ui}$,    $\ten{\hat{P}}_{ch}$ is the vector of vertices $\hat{P}_{ch_{j}}$ for the microstructural chain, and $\ten{\tilde P}_{1}(  \ten{\hat{\lambda}}_{u})  $ are the Piola stress predictions for the macroscopic test for the given uniaxial stretch values.  Then, we minimize the
quadratic error between the values obtained from experimental points,
$\ten{P}_{1}(  \ten{\hat\lambda}_{u})  $, and the estimation
computed from the B-spline of the stress in the chains, $\ten{\tilde P}_{1}(
\ten{\hat{\lambda}}_{u})  $. The function is
\begin{equation}
g(  \ten{\hat{P}}_{ch})  =\frac{1}{2}\left[  \ten{\tilde P}_{1}%
(  \ten{\hat{\lambda}}_{1})  -\ten{P}_{1}(
\ten{\hat{\lambda}}_{1})  \right]^{T}\ten{W}\left[
\ten{\tilde P}_{1}(  \ten{\hat{\lambda}}_{1})  -\ten{P}%
_{1}(  \ten{\hat{\lambda}}_{1})  \right]
\end{equation}
where the design variables, to be obtained, are $\ten{\hat{P}}_{ch}$. In the above expression we included a diagonal matrix $\ten{W}$ to weight, if needed, the errors of the different experimental data; see details in \cite{latorre2018experimental} for this type of weights. 

Experimental data is usually not smooth, but we consider that the model should give smooth stress-strain behavior. Then, a smoothing procedure should also be included \cite{latorre2018experimental}.  The smoothness of the curve can be controlled using a
penalization in the second derivative, and stability conditions (increasing load in the chain for increasing stretch) can be
guaranteed using a penalization in the first derivative in case it is
negative. There are many algorithms for spline smoothing (see \cite{weinert2013fast}). Since this discussion is not the purpose of the present manuscript, we just mention a simple effective procedure employing  approximate derivatives using finite differences of the vertices (i.e. a so-called P-splines approach, see \cite{eilers1996flexible}). These finite differences are
\begin{align}
D_{j}^{\left(  1\right)  }&=\frac{{\hat{P}}_{ch,j+1}-{\hat{P}}_{ch,j}}{h}=\frac{1}{h}\left(
{\hat{P}}_{ch,j+1}-{\hat{P}}_{ch,j}\right)
\\D_{j}^{\left(  2\right)  }&=\frac{D_{j+1}^{\left(  1\right)  }-D_{j}^{\left(  1\right)  }}{h}=\frac{1}{h^{2}}\left({\hat{P}}_{ch,j+2}-2{\hat{P}}_{ch,j+1}+{\hat{P}}_{ch,j}\right)
\\D_{j}^{\left(  3\right)  }&=\frac{D_{j+1}^{\left(  2\right)  }-D_{j}^{\left(  2\right)  }}{h}=\frac{1}{h^{3}}\left({\hat{P}}_{ch,j+3}-3{\hat{P}}_{ch,j+2}+3{\hat{P}}_{ch,j+1}-{\hat{P}}_{ch,j}\right)
\end{align}
where $h=\Delta\lambda_{ch}$ is the size of each interval in the spline. These finite differences are used to build the corresponding matrices $\ten{D}^{(k)}$ containing the coefficients of $\ten{\hat P}_{ch}$.  The penalty terms are added to build the function to be minimized
\begin{align}
\min_{\ten{\hat{P}}_{ch}} \{  \tfrac{1}{2}\ten{\hat{P}}_{ch}^{T}\ten{\bar N}^{T}
\ten{W\bar N\hat{P}}_{ch} 
&- \ten{\hat{P}}_{ch}^{T}\ten{\bar N}^{T}\ten{WP}
_{1}
+\tfrac{1}{2} \ten{\hat{P}}_{ch}^{T}\ten{D}^{(1)T}\ten{\Omega}^{(1)}
\ten{D}^{(1)}\ten{\hat{P}}_{ch}\nonumber \\
&+\tfrac{1}{2} \ten{\hat{P}}_{ch}^{T}\ten{D}
^{(2)T}\ten{\Omega}^{(2)}\ten{D}^{(2)}\ten{\hat{P}}_{ch}
+ \tfrac{1}{2}\ten{\hat{P}}_{ch}^{T}\ten{D}^{(3)T}\ten{\Omega}^{(3)}\ten{D}^{(3)}\ten{\hat{P}}_{ch}\}
\end{align}
where $\ten{\Omega}^{(k)}$ is a diagonal  weighting matrix to increase/decrease the smoothing in some parts, if desired. The final \textit{linear} system of equations to solve the problem is
\begin{equation}
\ten{A}{\ten{\hat P}_{ch}}=\ten{b}\label{final system equations}
\end{equation}
where
\begin{align}
    \ten{A}&=\ten{\bar N}^{T}\ten{W\bar N}+\ten{D}^{(1)T}\ten{\Omega}^{(1)}
\ten{D}^{(1)}+\ten{D}^{(2)T}\ten{\Omega}^{(2)}
\ten{D}^{(2)}+\ten{D}^{(3)T}\ten{\Omega}^{(3)}
\ten{D}^{(3)} \\\ten{b}&=\ten{\bar N}^{T}\ten{WP}
_{1} 
\end{align}
Smoothing parameters may guarantee, when needed, smoothness and stability conditions directly imposed on $P_{ch}\left(\lambda_{ch} \right) $. The smoothing parameters may take default values, be prescribed by the user, or automatically computed. Some techniques for automatic spline smoothing are also widely known, see e.g. \cite{o1986automatic,eubank1999nonparametric}, but are not relevant to the discussion in this paper, so we do not elaborate further.

\section{Constitutive manifolds}
In this section we briefly address how to build the constitutive manifolds for the model and how to use them to produce simulations with similar efficiency as phenomenological models.
\subsection{Building pre-computed manifolds from the computed chain behavior} 

Once the spline function $P_{ch}(\lambda_{ch})$ representing the behavior of the typical chain is known, the behavior of the continuum under a general deformation mode, characterized by $\lambda_1,\lambda_2,1/(\lambda_1\lambda_2)$ in an incompressible material, can be determined using Eq. (\ref{dPsidli}). In practice, constitutive manifolds for $\partial\Psi(\lambda_1,\lambda_2,\lambda_3)/\partial\lambda_i$ may be used for efficient evaluation in a finite element program, so computational times are similar to those employing analytical phenomenological models as, for example, Ogden's model, avoiding the time-consuming numerical integration over the sphere during the simulation. These manifolds are constructed evaluating the energy derivatives for a grid of values $\{\hat\lambda_1,\hat\lambda_2\}\in[\lambda_{1min},\lambda_{1max}]\times[\lambda_{2min},\lambda_{2max}] $ and fitting, for example, a bidimensional spline. Any suitable bi-dimensional interpolation or approximation scheme may be employed (for example the simplest, but non-smooth piecewise bi-linear interpolation); we just briefly summarize a periodic B-spline smooth surface approach, based on periodic unidimensional B-splines.

\subsubsection{One-dimensional periodic cubic splines\label{1Dspline}}

Periodic B-splines are just a convenient representation of the equivalent periodic cubic splines, the latter defined from nodal values and the former defined from vertices. Cubic splines approximates a function $f(x)$ within a given interval $j$ by a polynomial
which may be written as\begin{equation}f(x)\simeq f_j(x\in[x_j,x_{j+1}]) =\underbrace{a_jx^3+b_jx^2+c_jx+d_j}_{\text{cubic spline arrangement}}= \underbrace{\sum_{i=0}^{3} N_{i+1}(x) \hat f_{j+i}}_{\text{B-spline arrangement}}\end{equation} where $\{a_j,...,d_j\}$ are coefficients computed from continuity of the function and its first and second derivatives (in the typical spline arrangement), and  $N_{i+1} $ are the rearranged cubic interpolation polynomials in B-spline form which factors $\hat f_{j+i} $ are the B-spline control vertices. Periodic B-splines, employ equal size intervals $\Delta x:=x_{j+1}-x_j$ with normalized coordinates $\xi(x\in[x_j,x_{j+1}])=(x-x_j)/\Delta x\,\in[0,1]$ so the $N_{i+1}(\xi)$ functions are the same for all intervals, allowing for the nice compact representation---note the abuse of notation to avoid proliferation of symbols; we leave the function dependencies explicit when needed to avoid confusion

\begin{equation}
f_j(\xi)=\overbrace{\underbrace{\left[\xi^3, \xi^2,\xi,1 \right]}_{\text{monomial basis }}\underbrace{\frac{1}{6}\left[\begin{array}{cccc}
-1 & 3 & -3 & 1 \\
3 & -6 & 3 & 0 \\
-3 & 0 & 3 & 0 \\
1 & 4 & 1 & 0 \\
\end{array} \right]}_\text{coefficients of monomials} }^{\text{local shape functions } [N_{1},...,N_4]  \text{ evaluated at }\xi}\,\,\,\, \overbrace{\left[\begin{array}{c}
\hat f_{j} \\
\hat f_{j+1} \\
\hat f_{j+2} \\
\hat f_{j+3} \\
\end{array}\right]}^{\text{B-spline vertices}}
=:\mathbf{\Xi}\mathbf{C}_N\mathbf{\hat f}_j \label{eq1dperiodic}\end{equation}     

Then, the approximation of $f(x^*)$ for the value $x^* $ is obtained first locating the interval $j$, e.g. by $j=\text{int}[(x^*-x_1)/ \Delta x]+1$, where int() takes the integer part, second by obtaining the normalized coordinate in such interval by $\xi^*=(x^*-x_j)/\Delta x $, and finally using Eq. (\ref{eq1dperiodic}) to obtain $f_j(\xi^*(x^*))\simeq f(x^*)$. Several $k=1,...$ values can be simultaneously obtained using the global vector of vertices, namely
\begin{equation}\left[ \begin{array}{c}
\vdots \vspace{-1ex}  \\
\vdots \\
\tilde f_k \\
\vdots  \\
\end{array}\right] = \left[ \begin{array}{cccccc}
 \hdots& \text{\tiny (column j)} & \hdots & \hdots & \hdots & \hdots  \\
 \hdots & \downarrow & \hdots & \hdots & \hdots & \hdots  \\
\text{\tiny\ (row k)}  \rightarrow & N_1(\xi^*) & N_2(\xi^*) & N_3(\xi^*) & N_4(\xi^*) &  \hdots  \\
  \hdots &  \hdots &  \hdots &  \hdots &  \hdots &  \hdots  \\
\end{array}\right]\begin{array}{cc}
\left[ \begin{array}{c}
\vdots \\
\hat f_{j} \\
\hat f_{j+1} \\
\hat f_{j+2} \\
\hat f_{j+3} \\
\vdots \\
\end{array} \right] \hspace{-3ex} &  \begin{array}{c}
 \\
\leftarrow \text{\tiny (row j)}\\
 \\
 \\
 \\
 \\
\end{array}
\end{array}\label{Eq 1Dmtx}\end{equation}where $\tilde f_k \equiv f_j(x^*_k)$ is the global notation for the B-spline approximation of $f(x^*_k)$. In matrix notation we write $\mathbf{\tilde f}=\mathbf{ N}\mathbf{\hat f}$. A practical property of the B-spline representation, used by the finite differences P-splines approach, is that the convexity of the hull of vertices is inherited by the spline, so smoothness properties are simpler to implement.

Of course, the evaluation of the derivative of a spline is also immediate using Eq. (\ref{eq1dperiodic}) and taking the derivative of the monomials vector \begin{equation} 
\frac{d\tilde f}{dx}=\frac{df_j}{d\xi}\frac{d\xi}{dx}=\frac{1}{\Delta x}\frac{d\mathbf{\Xi}}{d\xi}\mathbf{C}_N\mathbf{\hat f}_j=\frac{1}{\Delta x}[3\xi^2,2\xi,1,0]\mathbf{C}_N\mathbf{\hat f}_j 
\label{derivativespline}\end{equation} \subsubsection{Periodic bi-cubic  B-spline surfaces}

The 1D approach may be generalized to more dimensions; two in our case. Considering a function $f(x,y)$, the bidimensional B-spline approximation $f_{st}$ within a given patch $st$ is

\begin{equation}f(x,y)\simeq f_{st}(\xi(x),\upsilon(y))=\sum_{i=0}^{3}\sum_{j=0}^{3}\underbrace{N_{i+1}(\xi)N_{j+1}(\upsilon)}_{M_{i+1,j+1}(\xi,\upsilon)}\hat f_{s+i,t+j}\end{equation} The coordinates $x,y$ have been again normalized to $\xi,\upsilon$ in the applicable patch $st$. Equation (\ref{eq1dperiodic}) converts to the bi-dimensional patch as
\begin{equation} f_{st}(\xi,\upsilon)=\mathbf{\Xi}\mathbf{C}_N\mathbf{\hat f}_{st}\mathbf{C}_N^T\mathbf{\Upsilon}^T
\label{2dspline}\end{equation}
where in a similar way $\mathbf{\Upsilon} :=[\upsilon^3,\upsilon^2,\upsilon,1]$, and $\mathbf{\hat f}_{st}$ is in this case the matrix containing the bi-dimensional grid of vertices for the patch $st$, i.e.
\begin{equation}\mathbf{\hat f}_{st}=\left[ \begin{array}{cccc}
\hat f_{s,t} & \hat f_{s,t+1} & \hat f_{s,t+2} & \hat f_{s,t+3} \\
\hat f_{s+1,t} & \hat f_{s+1,t+1} & \hat f_{s+1,t+2} & \hat f_{s+1,t+3} \\
\hat f_{s+2,t} & \hat f_{s+2,t+1} & \hat f_{s+2,t+2} & \hat f_{s+2,t+3} \\
\hat f_{s+3,t} & \hat f_{s+3,t+1} & \hat f_{s+3,t+2} & \hat f_{s+3,t+3} \\
\end{array}\right] \end{equation}However, Eq. (\ref{2dspline}) is not convenient for surface fitting using matrix notation, so it can be rearranged as \begin{equation} f_{st}(\xi^*(x^*),\upsilon^*(y^*))=\left[M_{1,1}\;M_{2,1}\;M_{3,1}\;M_{4,1}\;M_{1,2}\;\hdots M_{4,4}\right]\left[\begin{array}{c}
\hat f_{s,t} \\
\hat f_{s+1,t} \\
\vdots \\
\hat f_{s+3,t+3} \\
\end{array}\right]\end{equation}
with $M_{i,j}(\xi^*,\upsilon^*)=N_i(\xi^*)N_j(\upsilon^*)$. For a family of values $\{x^*_k,y^*_k\},k=1,...$, we can write a similar equation to that of the 1D case, Eq. (\ref{Eq 1Dmtx}); i.e.  \begin{equation}\begin{array}{c}    \left[\begin{array}{c}

\vdots \\  \tilde f_k \\ \vdots 
\end{array}\right] \end{array}=\left[\begin{array}{cccccccc}
\hdots &\hdots  & \hdots & \hdots & \hdots & \hdots  & \hdots & \hdots\\
\hdots & M_{1,1} &  M_{2,1} & \hdots &  M_{1,2} & \hdots & M_{1,4} & \hdots \\
\hdots & \hdots & \hdots & \hdots & \hdots & \hdots & \hdots & \hdots \\
\end{array} \right]\left[\begin{array}{c}
\vdots \\
\hat f_{s,t} \\
\hat f_{s+1,t} \\
\vdots \\
\hat f_{s,t+1} \\
\vdots \\
\hat f_{s,t+4} \\
\vdots \\
\end{array} \right]\label{2D big}\end{equation}
with $\tilde f_k \equiv f_{st}(x^*_k,y^*_k)$ and where a box-assembling scheme similar to that of 2D finite elements \cite{bathe2006finite} needs to be employed (note that vertices in a patch are not all arranged continuously in the global vector, but usually by sweeping a coordinate). We do not elaborate more because this type of procedure is standard. Equation (\ref{2D big}) may be written in compact notation as $\mathbf{\tilde f}=\mathbf{M}\mathbf{\hat f}$, where $\mathbf{\hat f}$ are the vertices of the hull defining the B-spline and $\mathbf{\tilde f}$ are evaluations. Obviously, the same scheme may be extended to tri-dimensional B-splines, and so on.
\subsubsection{Evaluation and generation of manifolds}
The previous scheme may be used to evaluate and generate
the constitutive manifolds. Since in our case there are only two independent variables, namely $\lambda_1,\lambda_2$ ---note that by incompressibility $\lambda_3=1/(\lambda_1\lambda_2)$--- , the bi-dimensional case may be employed.
For establishing the manifold, we first determine its computational domain, say $[ \lambda_{1\,min},\lambda_{1\,max}]\times[ \lambda_{2\,min},\lambda_{2\,max}]$ and the proper discretization of the domain, i.e. $\Delta\lambda_1,\Delta\lambda_2$; outside this domain the solution is extrapolated. Then, with the values of the vertices $\hat P_{ch_i}$ obtained in the previous section, see Eq. (\ref{final system equations}), the values of the energy derivatives $\Psi_1:=\partial\Psi/\partial\lambda_1$ and $\Psi_2:=\partial\Psi/\partial\lambda_2$ are obtained via numerical integration in the sphere. Now, classical bi-dimensional splines may be used to generate the manifold. In the case of using the current B-splines approach, we are interested in the vertices (which may be substantially fewer than the previous data, in which case $\tilde\Psi_\alpha\simeq\Psi_\alpha, \alpha=1,2$). These are obtained as usual solving a linear system of equations of the type of Eq. (\ref{final system equations}), e.g. \begin{equation}\mathbf{A}_\Psi\mathbf{\hat\Psi}_\alpha=\mathbf{b}_\alpha\end{equation}  with $ \mathbf{A}_\Psi:=\mathbf{M}^T\mathbf{W}\mathbf{M}$ (plus smoothing terms if desired)
and $\mathbf{b}_\alpha:=\mathbf{M}^T\mathbf{W}\mathbf{\Psi}_\alpha$. With the knowledge of the vertices $\mathbf{\hat\Psi}_\alpha$ of the spline surface, for any given values of $\lambda_1,\lambda_2,\lambda_3=1/(\lambda_1\lambda_2)$, it is immediate to obtain the values of   $\tilde\Psi_\alpha\simeq\Psi_\alpha, \alpha=1,2$.
We note that upon the knowledge of one manifold, for example $\partial\Psi\left(\lambda_{1},\lambda_{2},\lambda_{3}\right)/\partial\lambda_{1}$, the rest of them may be determined by isotropy requirements, see Eqs. (\ref{rvec}) and (\ref{derivativesolution}),  e.g.

\begin{equation}
\dfrac{\partial\Psi\left(\lambda_{1}=\hat{\lambda}_{1},\lambda_{2}=\hat{\lambda}_{2},\lambda_{3}
=\hat{\lambda}_{3}\right)}{\partial\lambda_{1}}=\dfrac{\partial\Psi\left(\lambda_{1}=\hat{\lambda}_{2},\lambda_{2}
=\hat{\lambda}_{1},\lambda_{3}=\hat{\lambda}_{3}\right)}{\partial\lambda_{2}}
\end{equation}

Of course the manifolds may be constructed also for $P_1$ and $P_2$ directly with the plane stress condition already imposed. They may be constructed also in terms of the principal invariants of the Cauchy-Green deformation tensor $I_1,I_2$, replacing the stretches as variables.

\subsection{Stress tensor and constitutive tangent for finite element analysis}

The typical finite element implementation in Total Lagrangean formulation involves the second Piola-Kirchhoff (2ndPK) stress $\ten{S}$ and its derivative $\mathbb{C}=d\ten{S}/d\ten{A}$ respect to the Green-Lagrange strains $\ten{A}$. Aside, the incompressibility condition is substituted by a quasi-incompressible one by adding a volumetric stored energy term $\mathcal{U}(J)$, where $J=\lambda_1\lambda_2\lambda_3$ is the Jacobian of the deformation, i.e. the strain energy function is
\begin{equation} 
\Phi(\ten{A})=\mathcal{U}(J)+\Psi(\lambda_1^d,\lambda_2^d,\lambda_3^d)
\end{equation}where $\lambda_i^d:=J^{-\tfrac{1}{3}}\lambda_i$ are the isochoric stretches of the deformation gradient. The 2ndPK stress is obtained through the chain rule as---only explicit sums on repeated indices are considered
 \begin{equation} 
 \ten{S}=2\frac{d\Phi}{d\ten{C}}=\frac{d\Phi}{d\ten{A}} 
 = \frac{d\mathcal{U}}{dJ}\frac{dJ}{d\ten{A}}+\sum_{k=1}^{3}{\frac{\partial\Psi}{\partial\lambda_k^d}}\frac{\partial\lambda_k^d}{d\ten{A}}  =:\ten{S}^v+\ten{S}^d\label{eqS}
 \end{equation} where  $\Psi_j\equiv\partial\Psi/\partial\lambda_j^d$ are immediately obtained from the previously computed manifold by the simple evaluation of the approximant B-spline surface ---see Eq. (\ref{2dspline})--- for the given isochoric stretch values. The pressure $p=d\mathcal{U}/dJ$ is obtained from the selected penalty function. Consider
\begin{align} 
\frac{dJ}{d\lambda_i}&=\frac{J}{\lambda_i}\\
 \frac{d\lambda_k^d}{d\lambda_i}
&=J^{-\frac{1}{3}}\left(\delta_{ik}-\tfrac{1}{3}\frac{\lambda_k}{\lambda_i}\right)\label{dlkd}\\
\frac{d\lambda_i}{d\ten{A}}&=\frac{1}{\lambda_i}\ten{N}_i\otimes\ten{N}_i
\label{dlambda}\end{align}
where $\ten{N}_i$ are the eigenvectors of the spectral decompositions of $\ten{C}$ and $\ten{A}$.  Then, the kinematic second order tensors in Eq. (\ref{eqS}) are easily obtained using the spectral decomposition
\begin{equation} 
\ten{J}:=\frac{dJ}{d\ten{A}}=\sum_{i=1}^{3}\frac{dJ}{d\lambda_i}\frac{d\lambda_i}{d\ten{A}}\ten{N}_i\otimes \ten{N}_i=\sum_{i=1}^{3}\frac{J}{\lambda_i^2}\ten{N}_i\otimes \ten{N}_i=J\ten{C}^{-1}
\end{equation}
\begin{align}
\ten{\lambda}_k^d&:=\frac{d\lambda_k^d}{d\ten{A}}=\sum_{i=1}^3\frac{\partial\lambda_k^d}{\partial\lambda_i}\frac{d\lambda_i}{d\ten{A}} =\sum_{i=1}^{3}{\frac{J^{-\frac{1}{3}}}{\lambda_i}\left(\delta_{ik}-\tfrac{1}{3}\frac{\lambda_k}{\lambda_i}\right)}\ten{N}_i\otimes \ten{N}_i \label{Lambdak}
\end{align}
Therefore, the stress tensor is
\begin{equation}\ten{S}^v=\underbrace{\frac{d\mathcal{U}}{dJ}J\ten{C}^{-1}}_{\ten{S}^v}
+\underbrace{\sum_{i=1}^{3}\sum_{k=1}^{3}\frac{{\Psi_k}J^{-\frac{1}{3}}}{\lambda_i^2}\left(\lambda_i\delta_{ik}-\tfrac{1}{3}{\lambda_k}\right)\ten{N}_i\otimes \ten{N}_i}_{\ten{S}^d}\end{equation}

The computation of the tangent may also be performed by systematic use of the chain rule:
\begin{align}
\mathbb{C}&:=\frac{d\ten{S}}{d\ten{A}}=\frac{d^2\Phi}{d\ten{A}d\ten{A}}=\frac{d\ten{S}^v}{d\ten{A}}+\frac{d\ten{S}^d}{d\ten{A}}
\\&=\underbrace{\frac{d\mathcal{U}}{dJ}\frac{d^2J}{d\ten{A}d\ten{A}}+
\frac{d^2 \mathcal{U}}{dJ^2}\frac{dJ}{d\ten{A}}\otimes\frac{dJ}{d\ten{A}}}_{\mathbb{C}^v\,=\,{d\ten{S}^v}/{d\ten{A}}=\,d^2\mathcal{U}/(d\ten{A} d\ten{A})}+\underbrace{
\sum_{k=1}^{3}{\frac{\partial\Psi}{\partial\lambda_k^d}}\frac{d^2\lambda_k^d}{d\ten{A}d\ten{A}}
+\sum_{k=1}^{3}\sum_{p=1}^{3}  \frac{\partial^2\Psi}{\partial\lambda_k^d\partial\lambda_p^d}\frac{d\lambda_k^d}{d\ten{A}}\otimes\frac{d\lambda_p^d}{d\ten{A}}
}_{\mathbb{C}^d\,=\,{d\ten{S}^d}/{d\ten{A}}\,=\,d^2{\Psi}/(d\ten{A} d\ten{A})}\label{Cbychain}\end{align} 
where the bulk modulus $\kappa =\ dp/dJ=d^2\mathcal{U}/dJ^2$ is obtained from the selected penalty function and ${\partial^2\Psi}/{\partial\lambda_k^d\partial\lambda_p^d}$ is obtained from the derivative of the manifold, see Eq. (\ref{2dspline}), using the scheme of Eq. (\ref{derivativespline}), i.e. $d\mathbf{\Xi}/d\xi$ and $d\mathbf{\Upsilon}/d\upsilon$, where $\xi$ and $\upsilon$ are the normalized coordinates for $\lambda_1^d$ and $\lambda_2^d$. Other option is to derive the tangent  in terms of the Cauchy-Green deformation tensor. Further details about the derivation of the tangent are given in \ref{ApC}

\section{Examples}

\subsection{Predictions for the Kawabata et al  \cite{kawabata1981experimental} experiments\label{SecKawabata1}}

One of the most challenging experimental sets on rubber-like materials are the Kawabata biaxial experiments. Kawabata performed biaxial tests on vulcanized rubber containing 8 phr sulfur. In these biaxial tests, they prescribed different fixed stretches $\lambda_1$ in one direction, varying the stretch $\lambda_2$ in the other in-plane direction. During each test, they measured the nominal stresses $P_1$ and $P_2$ in both directions, hence obtaining two families of curves, namely $P_1(\lambda_1,\lambda_2)$ and $P_2(\lambda_1,\lambda_2)$. Since the material behaves in a quasi-incompressible manner, and hence any deformation mode may be characterized with only two independent variables (e.g. either $\lambda_1,\lambda_2$ or $I_1,I_2$), the Kawabata families of curves represent the behavior of the material in any deformation mode in the range of the prescribed stretches. 

Our procedure is a macro-micro-macro procedure, meaning that from the macroscopically observed behavior during a test, using the known material structure, we reverse-engineer the behavior of the microstructural components that results in the prescribed macroscopic behavior. Then, since
that micro-structural behavior fully describes the macroscopic behavior, it is used to obtain the continuum behavior under any other deformation mode. The goodness of the obtained microstructural behavior is demonstrated by the accuracy in predicting any deformation mode, even though the former has been characterized from a single experimental curve.

\begin{figure}[htbp!]
\centering

\begingroup
\tikzset{every picture/.style={scale=1.}}
\begin{figure}[H]
\centering
\input{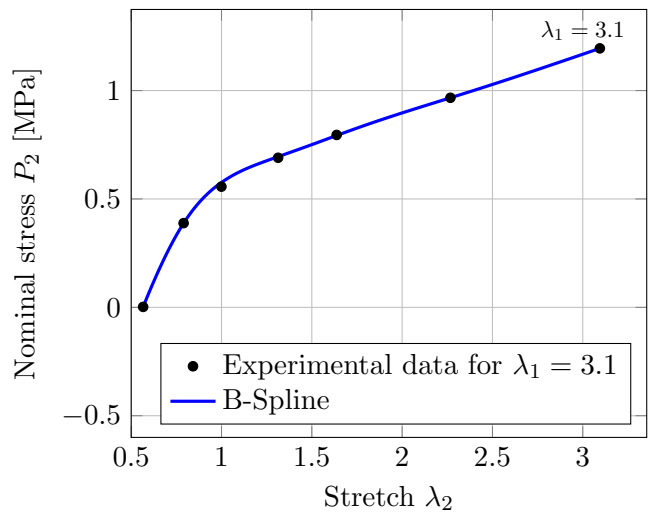}
\caption{Experimental data employed to compute $P_{ch}(\lambda_{ch})$ for the Kawabata et al material. Black points are the experimental data for $\lambda_1 = 3.1$. The continuous curve is a B-spline capturing those points. \label{F1K}}
\label{P2kawabataspline.tikz}
\end{figure}
\endgroup

\end{figure}

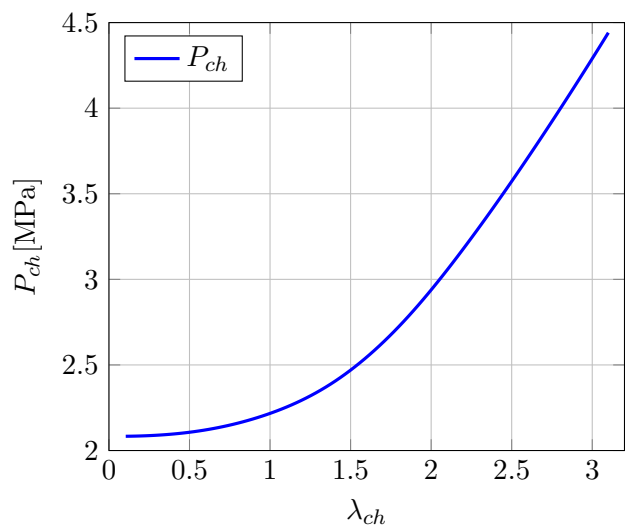
\begin{figure}[htbp!]
\centering

\begingroup
\tikzset{every picture/.style={scale=1.}}
\begin{figure}[H]
\centering
\begin{tikzpicture}
\begin{axis}[xmin = 0., ymin = 2,xmax = 3.2,ymax = 4.5,legend pos={north west}, xmajorgrids, ymajorgrids, xlabel={${\lambda}_{ch}$}, ylabel={${P}_{ch}$[MPa]}]
    \addplot[line width={1.2pt}, color={blue}, no markers]
        coordinates {
            (0.10423909455274577,2.0832673103326593)
            (0.13449930571887966,2.0836783507587526)
            (0.16475951688501353,2.084275097162577)
            (0.19501972805114742,2.0850692561593265)
            (0.2252799392172813,2.0860725343641993)
            (0.25554015038341515,2.08729663839239)
            (0.2858003615495491,2.0887532748590933)
            (0.31606057271568294,2.0904541503795064)
            (0.3463207838818168,2.0924109715688237)
            (0.37658099504795073,2.094635445042241)
            (0.4068412062140846,2.0971392774149544)
            (0.43710141738021846,2.0999341753021588)
            (0.4673616285463524,2.103031845319051)
            (0.49762183971248625,2.106443994080825)
            (0.5278820508786202,2.1101823282026775)
            (0.558142262044754,2.1142580885632527)
            (0.5884024732108879,2.1186796418429403)
            (0.6186626843770218,2.123454259378759)
            (0.6489228955431556,2.128589210351541)
            (0.6791831067092895,2.1340917639421164)
            (0.7094433178754234,2.139969189331317)
            (0.7397035290415573,2.146228755699975)
            (0.7699637402076912,2.15287773222892)
            (0.8002239513738251,2.1599233880989845)
            (0.8304841625399589,2.1673729924909995)
            (0.8607443737060928,2.175233814585797)
            (0.8910045848722267,2.1835131235642073)
            (0.9212647960383605,2.192218188607062)
            (0.9515250072044945,2.201356278895193)
            (0.9817852183706284,2.210935915525217)
            (1.0120454295367622,2.2209779184144485)
            (1.042305640702896,2.231510088162627)
            (1.07256585186903,2.242560305492106)
            (1.1028260630351638,2.2541564511252363)
            (1.1330862742012977,2.26632640578437)
            (1.1633464853674316,2.27909805019186)
            (1.1936066965335654,2.292499265070056)
            (1.2238669076996993,2.3065579311413114)
            (1.2541271188658334,2.3213019291279786)
            (1.2843873300319673,2.3367591397524077)
            (1.3146475411981011,2.352957443736952)
            (1.344907752364235,2.369924721803963)
            (1.3751679635303689,2.387688854675793)
            (1.4054281746965027,2.4062768152885314)
            (1.4356883858626366,2.4257003286059176)
            (1.4659485970287705,2.4459584673206742)
            (1.4962088081949043,2.4670499211531967)
            (1.5264690193610382,2.48897337982388)
            (1.556729230527172,2.5117275330531204)
            (1.586989441693306,2.5353110705613107)
            (1.6172496528594398,2.5597226820688483)
            (1.6475098640255736,2.584961057296128)
            (1.6777700751917077,2.6110248859635443)
            (1.7080302863578416,2.637912857791492)
            (1.7382904975239755,2.665623662500367)
            (1.7685507086901093,2.694155989810565)
            (1.7988109198562432,2.7235085294424795)
            (1.829071131022377,2.7536795609873526)
            (1.859331342188511,2.784653814584381)
            (1.8895915533546448,2.816399691156447)
            (1.9198517645207787,2.8488846194684414)
            (1.9501119756869125,2.8820760282852556)
            (1.9803721868530464,2.9159413463717785)
            (2.0106323980191805,2.9504480024929003)
            (2.040892609185314,2.98556342541351)
            (2.071152820351448,3.0212550438985013)
            (2.101413031517582,3.057490286712761)
            (2.131673242683716,3.0942365826211806)
            (2.1619334538498496,3.13146136038865)
            (2.1921936650159837,3.16913204878006)
            (2.2224538761821173,3.207216076560299)
            (2.2527140873482514,3.245680967789959)
            (2.2829742985143855,3.2845025491674384)
            (2.313234509680519,3.323671275281001)
            (2.3434947208466532,3.3631790897142104)
            (2.373754932012787,3.403017936050631)
            (2.404015143178921,3.443179757873828)
            (2.4342753543450546,3.4836564987673655)
            (2.4645355655111887,3.5244401023148098)
            (2.4947957766773223,3.5655225120997236)
            (2.5250559878434564,3.6068956717056726)
            (2.55531619900959,3.6485515247162215)
            (2.585576410175724,3.690482014714935)
            (2.615836621341858,3.732679085285377)
            (2.646096832507992,3.775134680011114)
            (2.6763570436741255,3.817840746008655)
            (2.7066172548402596,3.860791025131775)
            (2.7368774660063937,3.9039839686530757)
            (2.7671376771725273,3.9474187909617875)
            (2.7973978883386614,3.9910947064471474)
            (2.827658099504795,4.035010929498384)
            (2.857918310670929,4.079166674504734)
            (2.888178521837063,4.123561155855428)
            (2.918438733003197,4.168193587939699)
            (2.9486989441693305,4.213063185146781)
            (2.9789591553354646,4.258169161865906)
            (3.0092193665015983,4.303510732486307)
            (3.0394795776677324,4.3490871113972185)
            (3.069739788833866,4.39489751298787)
            (3.1,4.440941151647499)
        }
        ;
    \addlegendentry {${P}_{ch}$}
\end{axis}
\end{tikzpicture}
\caption{$P_{ch}(\lambda_{ch})$ obtained from Kawabata's experimental data  $P_2(\lambda_2)$ with $\lambda_1=3.1$  \label{PchKawa}}
\label{Pchkawabata.tikz}
\end{figure}
\endgroup

\end{figure}

\begin{figure}[htbp!]
\centering
\includegraphics{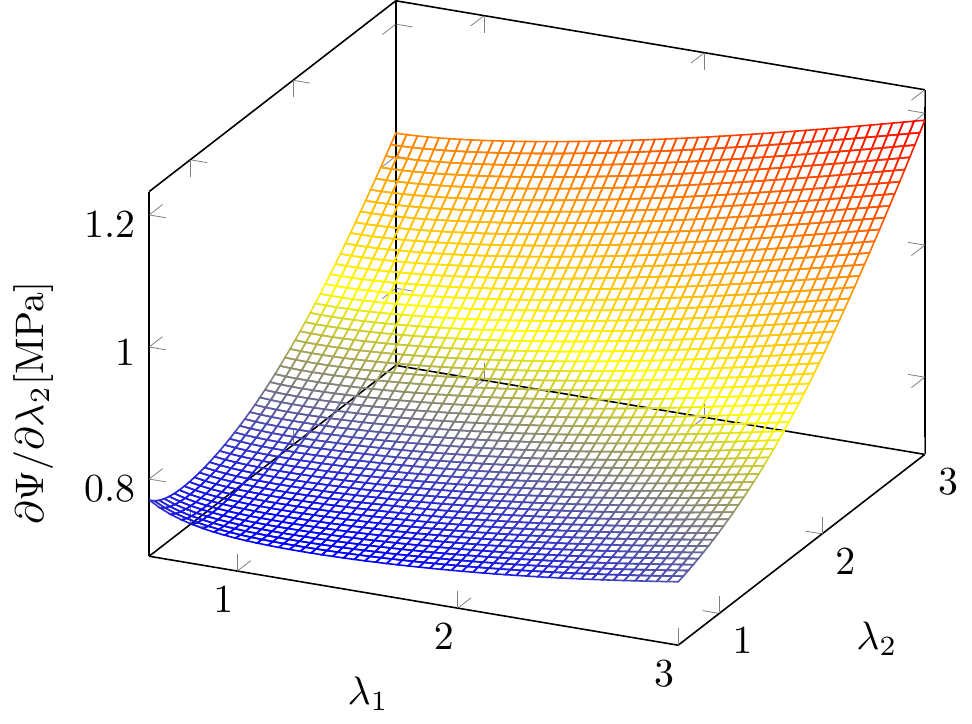}
\caption{Constitutive manifold obtained from Kawabata's curve  $P_2(\lambda_2)$ with $\lambda_1=3.1$  \label{manifold} }
\end{figure}

Then, to compute the representative micro-structural behavior of Kawabata's material, \emph{we will just use one of the stress-strain curves from those tests}, in particular the $\{\lambda_2,P_2\}$ points for  $\lambda_1=3.1$, see Fig. \ref{F1K}. Using the above procedure, solving the linear system of equations, we obtain the $\hat P_{ch}$ values and then we build a B-spline using those points to have the continuous function $P_{ch}(\lambda_{ch})$ which represents the behavior of the typical chain of the material; this function is shown in Fig. \ref{PchKawa}. With that function, we can compute the behavior of the material for any arbitrary deformation mode. For instance, we can compute the constitutive manifold $\partial\Psi(\lambda_1,\lambda_2,\lambda_3=(\lambda_1\lambda_2)^{-1})/\partial\lambda_1$, which is shown in Fig. \ref{manifold}. 

\begin{figure}[htbp!]
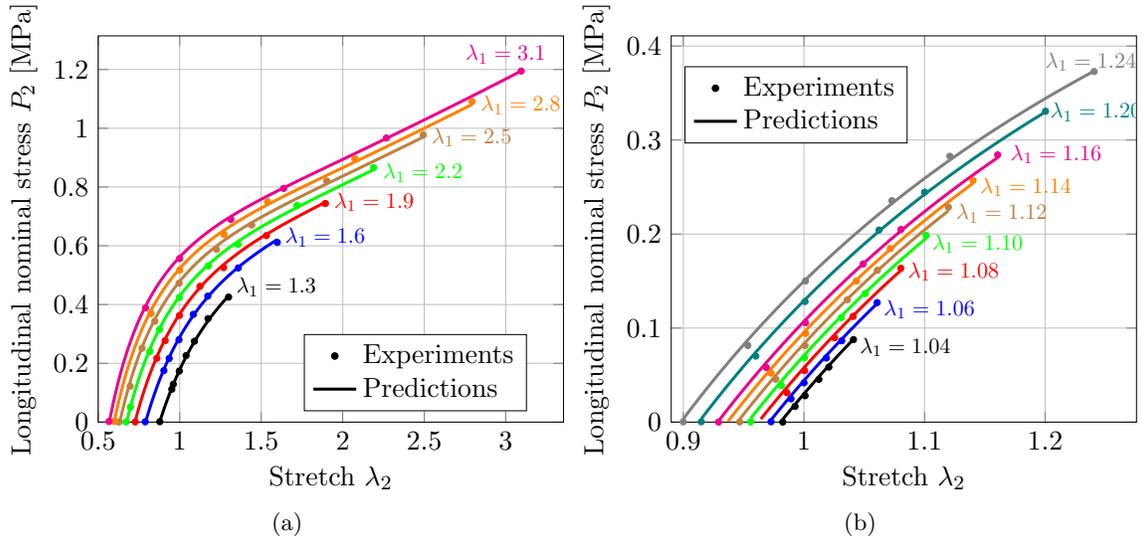


\begingroup
\tikzset{every picture/.style={scale=0.95}}
\begin{figure}[H]
\centering
\subfloat[]{\input{P2kawabatafirstrank.tikz}}
\subfloat[]{\input{P2kawabatasecondrank.tikz}}
\caption{Kawabata et al \cite{kawabata1981experimental} experiments. Predictions for the longitudinal nominal stress $P_2(\lambda_2)$  obtained with the present model for the range $\lambda_1$ from (a) $\lambda_1=1.3$ to $\lambda_1=3.1$ and (b)  $\lambda_1=1.04$ to $\lambda_1=1.24$ \label{KP2l2}}
\label{P2kawabatafirstrank.tikz}
\end{figure}
\endgroup

\end{figure}

\begin{figure}[htbp!]
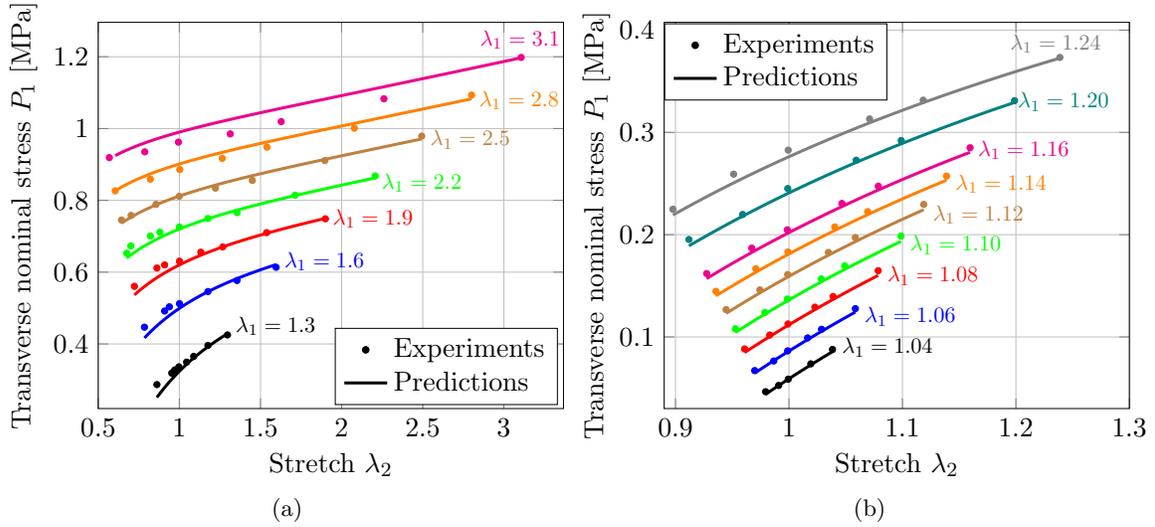


\begingroup
\tikzset{every picture/.style={scale=0.95}}
\begin{figure}[H]
\centering
\subfloat[]{\input{P1kawabatafirstrank.tikz}}
\subfloat[]{\input{P1kawabatasecondrank.tikz}}
\caption{Kawabata et al \cite{kawabata1981experimental} experiments. Predictions for $P_1(\lambda_2)$ obtained with the present model for the range $\lambda_1$ from (a) $\lambda_1=1.3$ to $\lambda_1=3.1$ and (b)  $\lambda_1=1.04$ to $\lambda_1=1.24$ \label{KP1l2}}
\label{P1kawabatafirstrank.tikz}
\end{figure}
\endgroup

\end{figure}

Using either these manifolds or integrating numerically, we can compute the behavior of the material for  the different combinations of $\lambda_1,\lambda_2$ in the Kawabata experiments, and hence obtain predictions of the $P_1(\lambda_1,\lambda_2),P_2(\lambda_1,\lambda_2)$ experimental data. Note that we can also build the manifolds $P_1(\lambda_1,\lambda_2)$ and $P_2(\lambda_1,\lambda_2)$ to obtain immediately the stress for any loading condition with the constrains of incompressible and plane stress already enforced. Since Kawabata's curves for the biaxial tests are plotted for varying $\lambda_2$ in the abscissae, for reference we label Axis-$2$ as longitudinal and Axis-$1$ as transverse. In Figure \ref{KP2l2} we show the comparison of our predictions for the longitudinal nominal stress $P_2(\lambda_2)$ with the experimental data from Kawabata et al for different fixed values of transverse stretch $\lambda_1$. It is seen that even though we used only one test curve to obtain $P_{ch}(\lambda_{ch})$, the predictions are very accurate for all the values of $\lambda_1$ and $\lambda_2$. Furthermore, in Figure \ref{KP1l2} we show the transverse nominal stresses for the same tests, i.e. $P_1(\lambda_2)$ for the same fixed values of transverse stretch  $\lambda_1$.  These curves show similar accuracy.

 To compute the material parameters of their model for the Kawabata et al material, Khi\^{e}m and Itskov used three curves (uniaxial, equibiaxial and pure shear) from Treloar's test. They obtained very good predictions for all sets of tests and noted that both materials are in essence the same material. However, following Urayama \cite{urayama2006experimentalist}, and because of the issues explained in \ref{Sec chain}, they noted the need of using all three tests from Treloar's experiments to calibrate their model, which contained terms in the two principal invariants. In practice, even being micro-mechanically motivated, their model contains two intrinsic degrees of freedom, meaning that it is defined by two different functions of two invariants, each one with its own material parameters, see Eq. (25) in Ref. \cite{khiem2016analytical}. In contrast, our model takes full advantage of the assumed structure of the material, as composed of fibers isotropically distributed. Remarkably, no further assumption is made, and the material behavior is fully characterized by a scalar function $P_{ch}(\lambda_{ch})$, so it has only one degree of freedom as the Arruda-Boyce model. Then, as we have done for the Kawabata material, the material behavior may be fully determined from a single test as in the Arruda-Boyce model. The only limitation is that to avoid extrapolations, the test must be performed to a sufficient range of stretch, so the obtained  $P_{ch}(\lambda_{ch})$ determines the behavior for all the desired deformations (i.e. these do not exceed the computed $\lambda_{ch}$ range). Of course extrapolations may be used appending, for example, a rational function with an estimated locking stretch. Furthermore, global analytical functions, instead of P-splines, could be used to fit the microstructural behavior data $\left\lbrace\lambda_{ch},P_{ch} \right\rbrace $

The Khi\^{e}m and Itskov predictions in Fig. 4 of Ref. \cite{khiem2016analytical} for the Kawabata et al experiments, using Treloar's curves for calibrating the model, are accurate, but our predictions for the Kawabata experiments in Fig. \ref{KP2l2}, using one of their curves to calibrate the model, are even more accurate. However, one of the reasons  could be that the material from Kawabata and the material from Treloar, having the same chemical composition have slightly different behavior. The study of this issue is addressed in \ref{appen_Tre_kawa}, where the function $P_{ch}(\lambda_{ch})$ from Treloar's tests is obtained and then applied to Kawabata's experiments.         
\subsection{Predictions for the Kawamura et al \cite{Kawamura} experiments }
Kawamura et al \cite{Kawamura} performed biaxial experiments on two Poly-dimethylsiloxane (PDMS) network materials. PDMS or dimethicone is a type of silicone used, for example, in contact lenses, microfluidic chips or medical devices, and has a siloxane backbone, so expected mechanical behavior is different from the typical carbon-based rubber \cite{Lewis_silicone}. The two tested cross-liked materials by Kawamura et al are obtained, respectively from melt PDMS and from a concentrated solution (70wt \%), with solvent being Trimethyl-terminated oligo-dimethylsiloxane; see details in \cite{Kawamura}. Their biaxial experiments covered all the deformation domain from uniaxial to equibiaxial stretchings. 

\begin{figure}[htbp!]
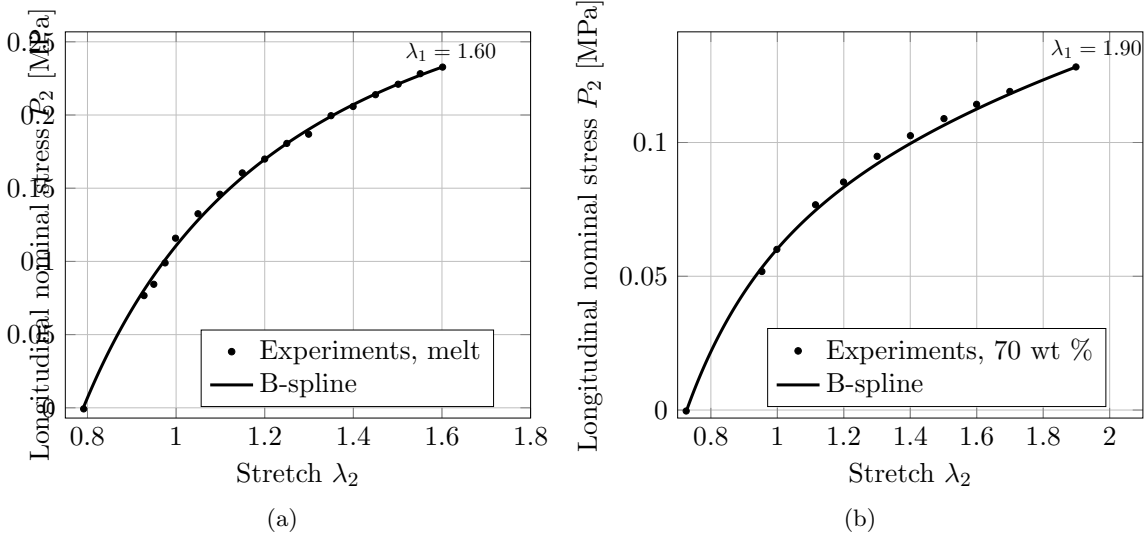


\begingroup
\tikzset{every picture/.style={scale=0.95}}
\begin{figure}[H]
\centering
\subfloat[]{\input{kawamuraP2_melt_usada.tikz}}
\subfloat[]{\input{kawamura_usada.tikz}}
\caption{Kawamura et al \cite{Kawamura} experiments for PDMS(melt and solution 70wt \%). Experimental data used to calibrate the material model and corresponding B-splines. (a) melt with $\lambda_1=1.6$, (b) 70wt \% solution with $\lambda_1=1.9$. \label{Kusadas}}
\label{kawamuraP2_melt_usada.tikz}
\end{figure}
\endgroup

\end{figure}

To perform the predictions for these tests we followed the same procedure as with the Kawabata et al experiments. First we select one curve from their experiments (the one which has the largest computational domain to avoid extrapolations). The selected experimental data are shown in Fig. \ref{Kusadas} along the B-spline used to obtain the respective $P_{ch}(\lambda_{ch})$ functions. Thereafter, with the obtained micromechanical behavior, we predict the rest of the experimental curves for both materials. The comparion for the melt material is shown in Fig. \ref{Kwml2}, whereas the comparison for the 70wt \% solution is shown in Fig. \ref{Kw70l2}. It can be observed that for both materials, both the longitudinal and the transverse stresses are predicted to excellent accuracy.

\begin{figure}[htbp!]

\begingroup
\tikzset{every picture/.style={scale=0.95}}
\begin{figure}[H]
\centering
\subfloat[]{\input{kawamuraP1_melt.tikz}}
\subfloat[]{\input{kawamuraP2_melt.tikz}}
\caption{Kawamura et al \cite{Kawamura} experiments for melt. Predictions for $P_1(\lambda_1,\lambda_2)$ (a) and $P_2(\lambda_1, \lambda_2)$ (b) obtained with the present model, compared with experimental data \label{Kwml2}}
\label{kawamuraP1_melt.tikz}
\end{figure}
\endgroup

\end{figure}

\begin{figure}[htbp!]

\begingroup
\tikzset{every picture/.style={scale=0.95}}
\begin{figure}[H]
\centering
\subfloat[]{\input{kawamuraP1.tikz}}
\subfloat[]{\input{kawamuraP2.tikz}}
\caption{Kawamura et al \cite{Kawamura} experiments for solution (70wt \%). Predictions for $P_1(\lambda_1,\lambda_2)$ (a) and $P_2(\lambda_1, \lambda_2)$ (b) obtained with the present model, compared with experimental data \label{Kw70l2}}
\label{kawamuraP1.tikz}
\end{figure}
\endgroup

\end{figure}
        
\section{Conclusions}        
In this paper we have introduced a new data-driven structure-based procedure to model isotropic, rubber-like materials. The approach is a macro-micro-macro approach in the sense that the micromechanical behavior of the fibers is obtained, including all the interactions, directly from macroscopic tests solving a linear system of equations. This micromechanical behavior may be subsequently employed to build macroscopic constitutive manifolds to efficiently predict the behavior of the continuum via finite elements under any loading condition. 
We have applied the approach to predict, to very good accuracy, the series of biaxial tests from Kawabata et al on vulcanized rubber containing 8-phr sulfur, and from Kawamura et al on two silicone materials. In these materials we used only one of the test curves to obtain the microstructural behavior of a representative chain. The predictions using this reverse-engineered microstructural behavior are excellent under any studied deformation mode.

We have also applied the procedure to obtain the behavior of a representative chain in Treloar's material, see \ref{appen_Tre_kawa}. This fiber behavior has subsequently been used to predict both Treloar's tests and Kawabata's biaxial test series. It is seen that whereas both materials are very similar, some differences may be appreciated in their behavior. 

As additional material, we include data and commented code in Julia language to reproduce the examples from Kawabata et al and from Treloar.      

The present approach is nonparametric, and micro-mechanical assumptions about the behavior of the material are reduced to a minimum. However, excellent accuracy is still obtained using one curve to calibrate the material because the microstructure of the material (isotropic distribution of chains, fibers or networks) is accounted for. Therefore, the macro-micro-macro approach looks promising for reverse-engineering other types of materials as, for example, biological tissues.    
\section*{Acknowledgements}

Partial financial support for this work has been given by grant
PGC-2018-097257-B-C32 from the Ministerio de Ciencia, Innovaci\'{o}n y
Universidades of Spain. 
\begin{appendix}

\section{About chain models and their pre-integration}\label{Sec chain}

The behavior of the chains in structure-based polymer models frequently follow statistical mechanics approaches. Initial Gaussian treatments evolved into more accurate Langevin statistical treatments to account for the locking stretch when the distance between chain ends $r_{ch}\rightarrow Nl$.\ These models require the evaluation of the inverse of the Langevin function\textbf{\ }
$\mathcal{L}^{-1}\left(  r_{ch}/Nl\right)  =y$, where $\mathcal{L}(y)=coth(y)-1/y$ is the Langevin function. 
Considering Langevin-based statistics, the probability distribution $\mathsf{p}(r_{ch})$
gives the following chain entropy \cite{treloar1975physics}

\begin{equation}
s\left(  r_{ch}\right)  =k\ln\mathsf{p}\left(  r_{ch}\right)  =k\left[
c-N\left(  \frac{r_{ch}}{Nl}y+\ln\frac{y}{\sinh y}\right)  \right]
\end{equation}where $k$ is the Boltzmann constant and $c$ is another constant. The stretch of the chain is ${\lambda}_{ch}=r_{ch}/$ $r_{0}=$
$r_{ch}/(\sqrt{N}l)$, which reaches a locking value of  ${\lambda}_{lock}=\sqrt{N}$.
Then, neglecting changes in internal energy as usual in these models, the tension in the chain  can be obtained from the
thermodynamic requirement

\begin{equation}
F_{ch}\left(  r_{ch}\right)  =-T\frac{ds\left(  r_{ch}\right)  }{dr_{ch}
}=\frac{kT}{l}\mathcal{L}^{-1}\left(  \frac{{\lambda}_{ch}}{\sqrt{N}
}\right)\label{Fch}
\end{equation}where $T$ is the absolute temperature, and the nominal stress is

\begin{equation}
P_{ch}\left(  r_{ch}\right)  =-T\frac{ds\left( {\lambda}_{ch}\right)
}{d{\lambda}_{ch}}=F_{ch}\left(  r_{ch}\right)  \frac{dr_{ch}}
{d{\lambda}_{ch}}=\sqrt{N}lF_{ch}\left(  r_{ch}\right)
\end{equation}For small values of $r_{ch}/Nl$, the Gaussian description is
recovered because by Taylor series

\begin{equation} 
\mathcal{L}^{-1}\left(  r_{ch}/Nl\right)  \simeq3\frac{r_{ch}}{Nl}=3\frac{{\lambda}_{ch}}{\sqrt{N}}+...
\end{equation} so $F_{ch}\left(  \lambda_{ch}\right)  =3l^{-1}kT\lambda_{ch}/\sqrt{N} $; for a more detailed explanation see \cite{amores2019ADES}. Note that for Langevin-based models, if  ${\lambda}_{ch}\rightarrow{\lambda}_{lock}$, then $\mathcal{L}
^{-1}\left(  {{\lambda}_{ch}}/{\sqrt{N}}\right)  \rightarrow
\mathcal{L}^{-1}\left(  1\right)  \rightarrow\infty$, so $P_{ch}\left(
\lambda_{ch}\right)  \rightarrow\infty$, whereas the Gaussian approach provides load values linear with the stretch. 

Models based on Langevin statistics are conceptually appealing, but have some limitations that should be beared in mind. The first one is the accurate evaluation of the inverse Langevin function itself. Since there is no known closed form, and accuracy in the evaluation is extremely important \cite{ammar2016effect}, several approximants have been proposed in the last years \cite{nguessong2014new,kroeger2015simple,itskov2012taylor,jedynak2017new,marchi2015error,darabi2015simple}. Comparisons are given in \cite{jedynak2017new} and \cite{benitez2018simple}. The most accurate approximant is given in \cite{benitez2018simple}, using a spline-based approach. A better alternative to the use of the Langevin function is given recently by Khi\^{e}m and Itskov \cite{khiem2016analytical} using the Rayleigh distribution function, resulting in a closed-form of the exact non-Gaussian probability distribution. In fact, the inverse Langevin function accurately represents the probability distribution only when $N$ is large enough (large chains, about $N=25\sim50$, see for example discussion around Figs. 6.3 and 6.6 in \cite{treloar1975physics}), but the values obtained from fitting to macroscopic tests may be near this value; see e.g. Figs 11 and 14 in \cite{arruda1993three}.
The second limitation is the purely entropic treatment of the stored energy, because it is well-known that stress-induced crystallization is relevant in some polymers at large strains, so the internal energy may have a considerable contribution; see for example Fig. 1.9 in \cite{treloar1975physics}, Fig. 3.5 of \cite{riande1999polymer}, Figs 3 and 4 in \cite{james1943theory} and references \cite{anthony1942equations,williams1971polymer}.

As mentioned, one of the best known models of this kind is the eight-chain model, which has only two parameters, namely $G=nkT$ (where $n$ is the chain density) and $N,$ easily obtained from a tensile test or other alternative test. With the information obtained from that test, relatively good predictions are obtained for other loading modes. The eight-chain model considers a microstructure of 8 chains, being the diagonals of a regular hexahedron oriented according to the principal directions of deformation. All 8 chains suffer the same microstructural (affine) stretch $\bar\lambda_{ch}$ under any continuum deformation. In Figure \ref{sphere}, the affine deformation process of a representative unit sphere of a polymeric material is depicted. In the initial configuration, at the left, the undeformed shape is spherical, turning into an ellipsoidal shape after the deformation, at the right. This kinematic process is mathematically characterized by the material deformation gradient tensor $\ten{X}$. $\ten{X}$ contains information about how the reference principal directions, $\ten{N}_{1}, \ten{N}_{2}$ and $\ten{N}_{3}$ are stretched and rotated during the deformation, turning out a new set of principal directions in the final configuration,  $\ten{n}_{1}, \ten{n}_{2}$ and $\ten{n}_{3}$. Therefore, the affine squared stretch $\lambda_{ch}^{2}$ of an arbitrary direction in spherical coordinates, $\ten{r}\left( \theta,\phi\right)$, can be obtained through $\ten{X}$ by $\lambda_{ch}^{2}=\left(  \ten{r\otimes r}\right)  \colon\ten{C}$, where $\ten{C}=\ten{X}^T\ten{X}$ is the right Green-Cauchy deformation tensor and 

\begin{equation}
\ten{r}=r_1\ten{N}_1+r_2\ten{N}_2+r_3\ten{N}_3=\sin\phi\cos\theta\ten{N}_{1}+\sin\phi\sin\theta\ten{N}
_{2}+\cos\phi\ten{N}_{3}
\end{equation}
If we integrate $\lambda_{ch}^2$ over a sphere $S$ we have, after some algebra \cite{amores2019ADES},
that the mean squared stretch is the same as the squared stretch in each one of the 8 chains of the Arruda-Boyce model:\begin{equation}
\frac{1}{S}{{\displaystyle\int\nolimits_{S}}
\lambda_{ch}^{2}dS}
=\frac{\lambda_{1}^{2}+\lambda_{2}^{2}+\lambda_{3}^{2}}%
{3}=\frac{I_1}{3}=\bar{\lambda}_{ch}^{2}%
\end{equation}Thus,  $\bar\lambda_{ch}=\sqrt{I_{1}/3}$, where $I_1$ is the first principal invariant of the Cauchy-Green deformation tensor $\ten{C}$. Since we have the functional dependence $\bar\lambda_{ch}(I_1)$, this function may be inserted in Eq. (\ref{Fch}) to obtain the derivative of the stored energy density of the continuum $d\Psi(I_1)/dI_1$ upon consideration of a chain density of $n$; see \cite{arruda1993three} for more details.
However, if we assume that the stored energy may be written in the form $\Psi(\lambda_1,\lambda_2,\lambda_3)=\Psi(I_1)=\Psi(\bar\lambda_{ch})$ --note the abuse of notation to avoid the proliferation of symbols--, a WYPiWYG approach may be adopted, so the prescribed data is exactly captured without the need of considering the statistical treatment. Furthermore, if the eight-chain structure is adopted, the average chain behavior can also be obtained without considering the probability distribution; see \cite{amores2019ADES}. Noteworthy, the prescribed data (from a single test) may be captured to any precision, but as with the Arruda-Boyce model, the model fails to accurately predict the behavior for all deformation modes. As explained in \cite{amores2019ADES},  the reason is that an eight-chain configuration considers only a single continuum variable (i.e. the first invariant), but the possible continuum deformation modes of an isotropic incompressible material have two degrees of freedom. For instance, the ratio between the $I_2$ invariant and the $I_1$ invariant changes substantially from the uniaxial and pure shear deformations to the equibiaxial deformations, see Fig. 11 in \cite{amores2019ADES}. Therefore an accurate constitutive model must include the two macroscopic degrees of freedom, even though the microscopic behavior in an isotropic material could be characterized by a single variable. 

Since the first invariant represents only the mean squared stretch and filters the influence of the deviations from that mean value, the second invariant, along with the couplings between both invariants, must be present in the higher order moments. Generally for any direction of the space we can write
the squared stretch as the addition of the mean value and the deviation from that mean value\begin{equation}
\lambda_{ch}^{2}=\bar{\lambda}_{ch}^{2}+\delta_{ch}\end{equation}
where $\delta_{ch}=\lambda_{ch}^{2}-\bar{\lambda
}_{ch}^{2}$ and by definition
\begin{equation}
{\displaystyle\int\nolimits_{S}}
\delta_{ch}dS=0\label{intdelta}\end{equation}
Thus, the quantity $\delta_{ch}$ represents the deviation. 
The diagonals of the cube in the eight-chain model are average directions of
the sphere in terms of $\lambda_{ch}^{2}$ (they happen to have always the mean squared stretch).\ It is important to note that
$\frac{1}{S}\int_{S}\Psi^{\ast}\left(  \lambda_{ch}^{2}\right)
dS\neq\Psi^{\ast}\left(  \bar{\lambda}_{ch}^{2}\right)  $ \ \ and \ $\frac
{1}{S}\int_{S}\Psi^{\ast}\left(  \delta_{ch}\right)  dS\neq
\Psi^{\ast}\left(  0\right)  $; i.e. the average energy or the average of the stresses are not those obtained from an average stretch measure. The accuracy of the results provided by the
eight-chain model in particular, and the relevance of using average strain measures in general, can be understood by the following Taylor expansion centered
on $\bar{\lambda}_{ch}^{2}$.

\begin{equation}\label{taylorexpan}
\Psi^{\ast}\left(  \lambda_{ch}^{2}\right)  =\Psi^{\ast}\left(  \bar{\lambda
}_{ch}^{2}+\delta_{ch}\right)  =\Psi^{\ast}\left(  \bar{\lambda}_{ch}^{2}\right)  +\left.  \dfrac{d\Psi^{\ast}\left(  x\right)  }{dx}\right\vert_{x=\bar{\lambda}_{ch}^{2}}\delta_{ch}+\frac{1}{2}\left.  \dfrac{d^{2}
\Psi^{\ast}\left(  x\right)  }{dx^{2}}\right\vert _{x=\bar{\lambda}_{ch}^{2}}\delta_{ch}+...
\end{equation}
This expression may be integrated in the sphere to give the continuum, macroscopic energy. Note that the quantities $\Psi^*(\bar\lambda^2_{ch})$, $\left. d\Psi/dx\right|_{\bar\lambda^2_{ch}}$, $\left. d^2\Psi/dx^2\right|_{\bar\lambda^2_{ch}}$ are constant within the integral. Then, using Eq. (\ref{intdelta}), after some lengthy but straightforward algebra, we arrive at

\begin{equation}
\frac{1}{S}\int_{S}\Psi^{\ast}\left(  \lambda_{ch}^{2}\right)
dS=\Psi^{\ast}\left(
\bar{\lambda}_{ch}^{2}\right)+\left. \dfrac{d\Psi^{\ast}\left(
x\right)  }{dx}\right\vert _{x=\bar{\lambda}_{ch}^{2}}\frac{1}{S}%
\underset{=0}{\underbrace{%
{\displaystyle\int_{S}}
\delta_{ch}dS}}+\frac{1}{2}\left. \dfrac{d^{2}\Psi^{\ast}\left(
x\right)  }{dx^{2}}\right\vert _{_{x=\bar{\lambda}_{ch}^{2}}}\frac{1}{S}%
{\displaystyle\int_{S}}
\delta_{ch}^{2}dS+...
\end{equation}
%
turning out
\begin{equation}
\frac{1}{S}\int_{S}\Psi^{\ast}\left(  \lambda_{ch}^{2}\right)
dS=\underset{\text{eight-chain model}}{\underbrace{\Psi^{\ast}\left(
\bar{\lambda}_{ch}^{2}\right)  }}+0+\underset{\text{terms not taken into
account by eight-chain model}}{\underbrace{\left. \dfrac{d^{2}\Psi^{\ast
}\left(  x\right)  }{dx^{2}}\right\vert _{_{x=\bar{\lambda}_{ch}^{2}}}\frac
{2}{45}I_{\delta2} +\frac{3}{45}\left. \dfrac
{d^{3}\Psi^{\ast}\left(  x\right)  }{dx^{3}}\right\vert _{_{x=\bar{\lambda}%
_{ch}^{2}}}I_{\delta3}+...}}
\label{preint}\end{equation}
$2/45 I_{\delta2}$ and $3/45 I_{\delta3}$ has been obtained after the integration of the third and fourth terms, respectively, of the expansion series as follows

\begin{align*}
\frac{1}{2}\frac{1}{S}\int_{S}\delta_{ch}^{2}dS=\int_{S}\left(\lambda_{ch}^{2}-\bar{\lambda}_{ch}^{2}\right)^{2}dS=\frac{2}{45}\left(I_{1}^{2}-3I_{2} \right)=\frac{2}{45}I_{\delta2}\\
\frac{1}{6}\frac{1}{S}\int_{S}\delta_{ch}^{3}dS=\int_{S}\left(\lambda_{ch}^{2}-\bar{\lambda}_{ch}^{2}\right)^{3}dS=\frac{3}{45}I_{\delta3}
\end{align*}
where $I_{\delta3}$ is an invariant function of $6^{th}$ order powers of the stretches.
Noteworthy, this is a pre-integrated expression in which the first term $\Psi^{\ast}\left(
\bar{\lambda}_{ch}^{2}\right)=\Psi^{\ast}\left(I_1\right)$--recall the abuse of notation-- contains the stored energy of an eight-chain model and the second term containing $d\Psi^*/dx$ vanishes identically. The first appearance of the second invariant is in the third term  of the series. Including this term, the pursued different relation between the uniaxial and the equibiaxial tests is obtained. However, the subsequent terms of the series become increasingly important away from the reference state, when the stretch $\lambda^2_{ch}$ of the fibers is not close to the mean value $\bar\lambda^2_{ch}$, i.e. when $\delta_{ch}\gg0$. This is expected in rubber-like solids under general deformations. If we insist in using an average strain and only the first term of the series, we are in fact embedding the information of the remaining terms in that single term  when fitting experiments. However, the relation between terms change considerably for different tests, so they should be included explicitly in computing the stored energy $\Psi= \frac
{1}{S}\int_{S}\Psi^{\ast}dS $. 

\begin{figure}[h!]

\begingroup
\tikzset{every picture/.style={scale=1.}}
\begin{figure}[H]
\centering
\input{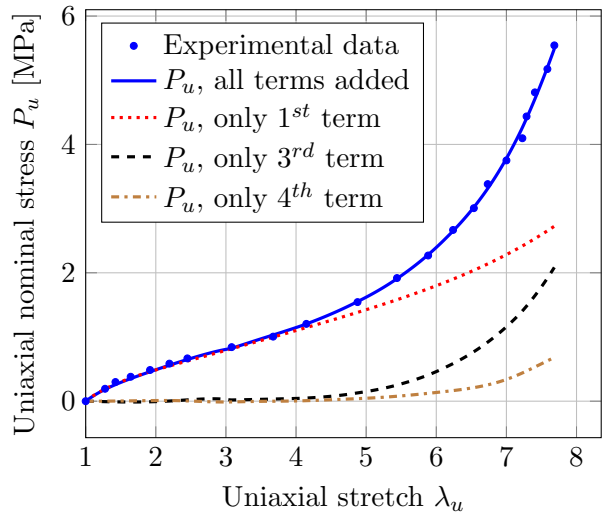}
\caption{Nominal stress, $P_{u}$ computed with a  WYPiWYG procedure including up to the fourth term of  Taylor's expansion of a pre-integrated scheme.\label{WYPIWYGpre}}
\label{tayloerexp.tikz}
\end{figure}
\endgroup

\end{figure}

A WYPiWYG procedure can be applied to  Eq. (\ref{preint}) to obtain the macroscopic stored energy $\Psi= \frac
{1}{S}\int_{S}\Psi^{\ast}dS $ by considering spline interpolations for $\Psi^*(x)$ and retaining several terms. In Fig.  \ref{WYPIWYGpre} we show such a computation retaining up to the $4^{th}$ term, and we plot the influence of each term in the sum. Of course, the procedure is capable of capturing the uniaxial test to high accuracy regardless of the number of terms considered. However, the higher order terms will have a different influence in other types of tests; note for example that the third term in  Eq. (\ref{preint}) includes the second invariant. The problem of such approach is that whereas
the high order terms vanish for small strains, at large strain may even become dominant, as the tendency in
 Fig.  \ref{WYPIWYGpre}  shows. In line with the number of series terms needed for an accurate evaluation of the inverse Langevin function \cite{itskov2012taylor}, unfortunately, the number of terms in the series of Eq. (\ref{preint}) that need to be considered is high, so such a procedure is expensive and the interpolation functions should be derivable as many times, complicating the procedure.


\section{Predictions for the Treloar material and for the Kawabata et al material using $P_{ch}(\lambda_{ch})$ obtained from Treloar's tests} \label{appen_Tre_kawa}
\subsection{$P_{ch}(\lambda_{ch})$ from the Treloar experiments}
Treloar made a well-known series of tests. The uniaxial tensile test, the equibiaxial test and the pure shear test are widely used to verify constitutive models of rubbers. In principle, just one test would be sufficient to characterize $P_{ch}(\lambda_{ch})$, and using this function, all tests should be predicted accurately. However, unfortunately the larger stretch reached during the tensile test does not cover all the domain in which  $P_{ch}(\lambda_{ch})$ is evaluated during the equibiaxial test, and vice-versa, the equibiaxial stretch values define a $P_{ch}(\lambda_{ch})$ domain which does not fully cover the domain needed for the tensile test by Treloar. Then, two options are possible.

\begin{figure}[htbp!]

\begingroup
\tikzset{every picture/.style={scale=1.}}
\begin{figure}[H]
\centering
\input{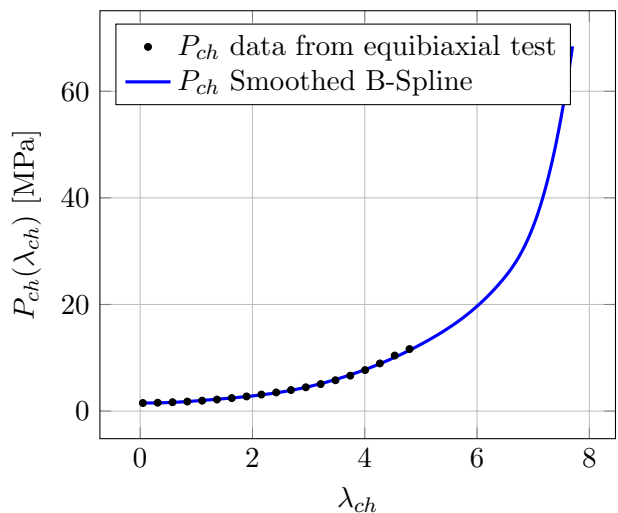}
\caption{$P_{ch}$ obtained from an equibiaxial test and extended with a rational function\label{Pchbiaxrat}}
\label{biaxpchsmoothed.tikz}
\end{figure}
\endgroup

\end{figure}

\begin{figure}[htbp!]
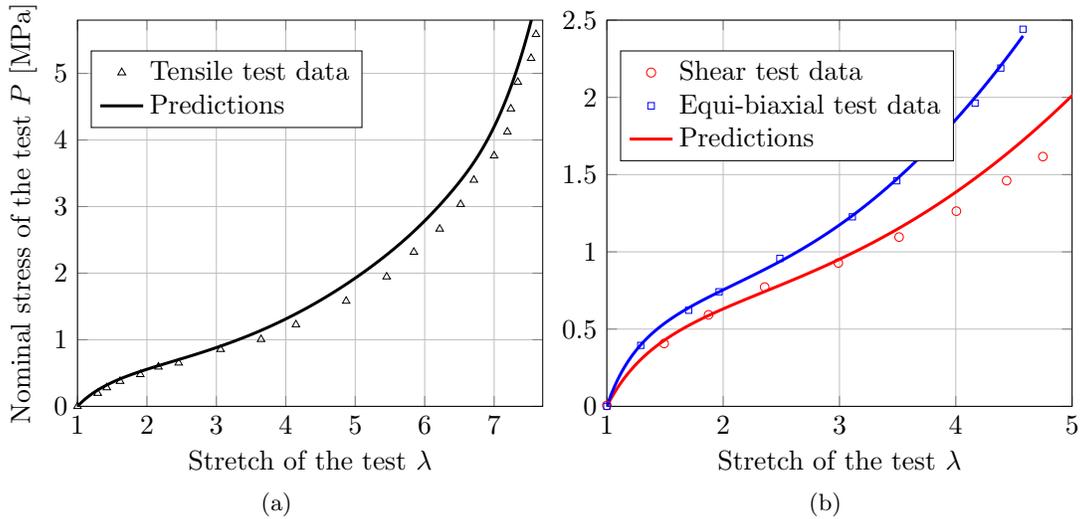


\begingroup
\tikzset{every picture/.style={scale=0.95}}
\begin{figure}[H]
\centering
\subfloat[]{\input{unipchfrombiaxaftercompleted.tikz}}
\subfloat[]{\input{unipchfrombiaxaftercompleted1.tikz}}
\caption{ Predictions of Treloar's tests using $P_{ch}$ captured from a biaxial test and extended with a rational function \label{Predrat}}
\label{unipchfrombiaxaftercompleted.tikz}
\end{figure}
\endgroup

\end{figure}

The first one is to insist in using just one test, for example the equibiaxial test, to characterize the material. Then, $P_{ch}(\lambda_{ch})$ must be extrapolated in the parts of the domain accessed  by the other tests but that were out of range during the equibiaxial test (or a global function may be fitted to the $P_{ch}(\lambda_{ch})$  data). Such approach is shown in Fig. \ref{Pchbiaxrat}, where the function $P_{ch}(\lambda_{ch})$ is extrapolated by a rational function of the type

\begin{equation}f(\lambda_{ch})=\frac{a\lambda_{ch}+b}{\lambda_{lock}^2-\lambda_{ch}^2}\end{equation}
to account for a limit stretch $\lambda_{lock}$ and a proper continuity with the experimental part given by the constants $a$ and $b$. The B-spline $P_{ch}(\lambda_{ch})$ function has been thereafter used to predict the three Treloar tests, shown in Fig. \ref{Predrat}, where we note that the equibiaxial test is predicted accurately, but the predictions of the uniaxial and pure shear tests have a larger error because the extrapolation employed is not directly based on experimental data. 

\begin{figure}[htbp!]

\begingroup
\tikzset{every picture/.style={scale=1.}}
\begin{figure}[H]
\centering
\input{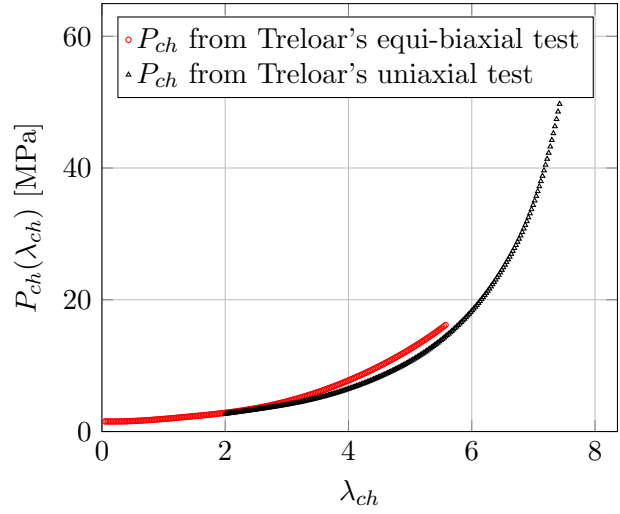}
\caption{  $P_{ch}$ data obtained from the uniaxial and equi-biaxial tests by Treloar and final adopted  B-spline curve. Note that the domain reached by both tests is different. Note also that in the overlapping part of the domain, the data obtained is very similar (although not identical). \label{Tre2test}}
\label{pchunipchbiaxjustknownpart.tikz}
\end{figure}
\endgroup

\end{figure}

\begin{figure}[htbp!]
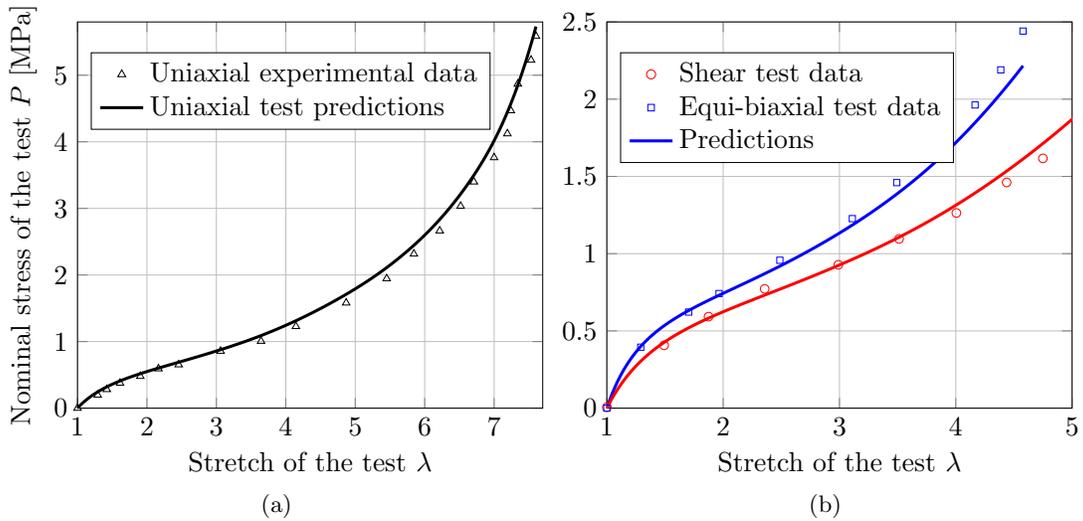


\begingroup
\tikzset{every picture/.style={scale=0.95}}
\begin{figure}[H]
\centering
\subfloat[]{\input{unipchf.tikz}}
\subfloat[]{\input{unipchf1.tikz}}
\caption{Predictions for Treloar's tests using a $P_{ch}(\lambda_{ch})$ obtained from tensile test data with the domain completed with some equibiaxial data\label{FinalTreloarpred}}
\label{unipchf.tikz}
\end{figure}
\endgroup

\end{figure}

A second option is to cover the domain using $P_{ch}(\lambda_{ch})$ data from two of Treloar's tests. This approach is shown in Fig. \ref{Tre2test}. In this figure it is observed that the domain reached by the equibiaxial test and by the uniaxial test are different, and that in the common domain, data is similar, although not identical. A regression smooth B-spline of all data from both tests is employed to create a $P_{ch}(\lambda_{ch})$ function that covers all the needed domain to predict Treloar's tests. These predictions are shown in Fig. \ref{FinalTreloarpred}, where it is clearly observed that in this case the equibiaxial test predictions are slightly less accurate, but the predictions for the other two tests have improved.

\subsection{Prediction of the Kawabata experiments using $P_{ch}(\lambda_{ch})$ from a Treloar material}

In Section \ref{SecKawabata1} we have shown that if the $P_{ch}(\lambda_{ch})$ function representing the behavior of a typical chain in the Kawabata material is known, we are able to build the constitutive manifolds that accurately predict the behavior of the material under any loading condition.
Following Khi\^{e}m and Itskov, we  will use the $P_{ch}(\lambda_{ch})$ function of Fig. \ref{Tre2test} to predict the Kawabata experiments to investigate to what extent both the Treloar and the Kawabata materials behave in a similar way.

\begin{figure}[htbp!]
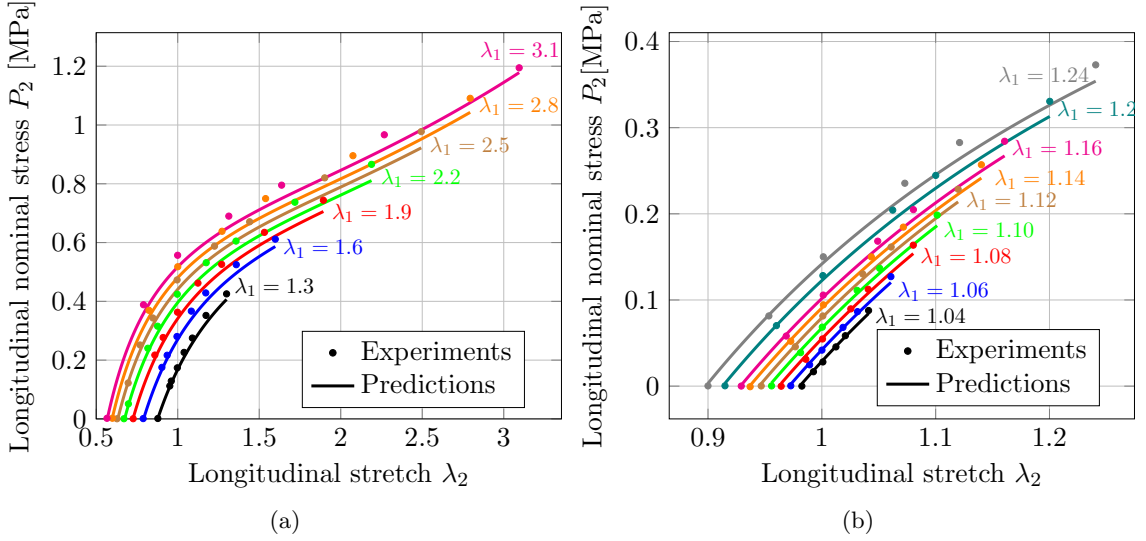


\begingroup
\tikzset{every picture/.style={scale=0.95}}
\begin{figure}[H]
\centering
\subfloat[]{\input{P2withtreloarfirstrank.tikz}}
\subfloat[]{\input{P2treloarsecondrank.tikz}}
\caption{Predictions for $P_2(\lambda_2)$ for the Kawabata experiments using $P_{ch}(\lambda_{ch})$ from  Treloar's material. Ranges of $\lambda_1$ (a) from $\lambda_1=1.3$ to $\lambda_1=3.1$ and (b)from $\lambda_1=1.04$ to $\lambda_1=1.24$ \label{KawafromTre1}}
\label{P2withtreloarfirstrank.tikz}
\end{figure}
\endgroup

\end{figure}

\begin{figure}[htbp!]
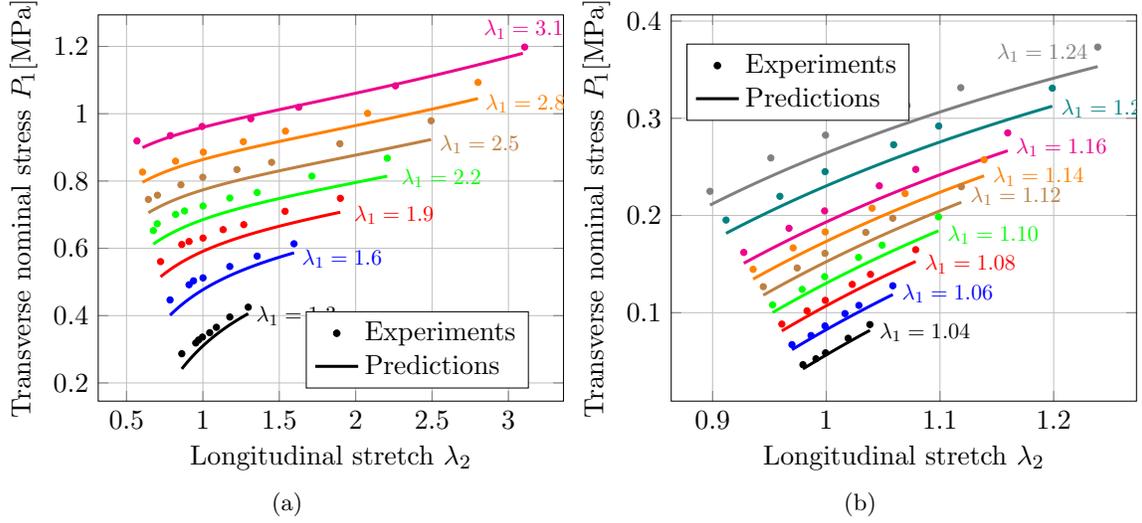


\begingroup
\tikzset{every picture/.style={scale=0.95}}
\begin{figure}[H]
\centering
\subfloat[]{\input{P1treloarfirstrank.tikz}}
\subfloat[]{\input{P1treloarsecondrank.tikz}}
\caption{Predictions for $P_1(\lambda_2)$ for the Kawabata experiments using $P_{ch}(\lambda_{ch})$ from  Treloar's material. Ranges of $\lambda_1$ (a) from $\lambda_1=1.3$ to $\lambda_1=3.1$ and (b)from $\lambda_1=1.04$ to $\lambda_1=1.24$ \label{KawafromTre2}}
\label{P1treloarfirstrank.tikz}
\end{figure}
\endgroup

\end{figure}

In Figs. \ref{KawafromTre1} and \ref{KawafromTre2} we show the predictions of the Kawabata experiments using Treloar's material. We show both the longitudinal nominal stress curves $P_2(\lambda_{2})$ and the transverse nominal stress curves $P_1(\lambda_2)$ for the different values of transverse stretch $\lambda_1$. These figures should be compared to Figs. \ref{KP2l2}. It is clearly observed that both materials are indeed similar, but not identical, because the present predictions are slightly worse than those shown in  Figs. \ref{KP2l2}. The small but appreciable differences are indeed manifest in transverse stresses $P_1(\lambda_2)$. We note that using $P_{ch}(\lambda_{ch})$ from Figs. \ref{Pchbiaxrat} and \ref{Tre2test} the same conclusion is obtained (differences using either curve are very small). 

\section{Derivation of the constitutive tangent \label{ApC}}
In this Appendix we derive the terms needed for the constitutive tangent in Eq. (\ref{Cbychain}). Taking $\Omega_{ij}$ as the components of the spin tensor of the eigenvectors in their own basis so $\ten{\dot N}_i=\sum_{j=1}^3 \Omega_{ji} \ten{N}_j$, the second derivative of the Jacobian is obtained from comparison of the rates
\begin{equation}{\ten{\dot J}}=\sum_{i=1}^{3}\left(\frac{J}{\lambda_i^3}\dot\lambda_i\right)\ten{N}_i\otimes \ten{N}_i+\sum_{i=1}^{3}\sum_{j\ne i}\left(\frac{J}{\lambda_j^2}-\frac{J}{\lambda_i^2}\right)\Omega_{ij}\ten{N}_i\otimes \ten{N}_j\end{equation}
\begin{align} 
\ten{\dot A}&=\sum_{i=1}^{3}\lambda_i\dot\lambda_i\ten{N}_i\otimes \ten{N}_i +\sum_{i=1}^{3}\sum_{j\ne i}\tfrac{1}{2}(\lambda_j^2-\lambda_i^2)\Omega_{ij}\ten{N}_i\otimes \ten{N}_j  \label{eqAdot}\end{align}
as
\begin{align}\frac{d^2J}{d\ten{A} d\ten{A}}&=\sum_{i=1}^{3}\frac{J}{\lambda_i^4}\ten{M}_{ii}\otimes\ten{M}_{ii}- \sum_{i=1}^{3}\sum_{j\ne i}\frac{2J}{\lambda_i^2\lambda^2_j} \ten{M}_{ij}\overset{s}{\otimes} \ten{M}_{ij}\nonumber\nonumber\\ &=J\ten{C}^{-1}\otimes\ten{C}^{-1}-2J\ten{C}^{-1}\overset{s}{\odot}\ten{C}^{-1}\end{align} where
\begin{align}
\ten{M}_{ij}&:=\ten{N}_i\otimes\ten{N}_{j}\\
\ten{M}_{ij}\overset{s}{\otimes}\ten{M}_{ij}&:=\tfrac{1}{4}\left[\ten{N}_i\otimes\ten{N}_{j}\otimes\ten{N}_i\otimes\ten{N}_{j}+\ten{N}_i\otimes\ten{N}_{j}\otimes\ten{N}_j\otimes\ten{N}_{i}\right. \nonumber\\ &\,\,\,\,\,\,\,+\left. \ten{N}_j\otimes\ten{N}_{i}\otimes\ten{N}_i\otimes\ten{N}_{j}+\ten{N}_j\otimes\ten{N}_{i}\otimes\ten{N}_j\otimes\ten{N}_{i} \right] \\
\left[\ten{C}^{-1}\overset{s}{\odot}\ten{C}^{-1}\right]_{ijkl}&:=\tfrac{1}{2}\left(C_{ik}^{-1}C_{jl}^{-1}+C_{il}^{-1}C_{jk}^{-1}\right) 
\end{align}
so
\begin{equation}
\mathbb{C}^v=\left(
J^2\frac{d^2 \mathcal{U}}{dJ^2}+J\frac{d\mathcal{U}}{dJ}\right)\ten{C}^{-1}\otimes \ten{C}^{-1}-2J\frac{d\mathcal{U}}{dJ}\ten{C}^{-1}\odot\ten{C}^{-1}
\end{equation}
In the derivative of the isochoric stress tensor $\ten{S}^d$, only the kinematic fourth order tensor $d\ten{\lambda}_k^d/d\ten{A}={d^2\lambda_k^d}/{d\ten{A}d\ten{A}}$, 
is still unknown. First consider
\begin{equation}
\ten{\dot\lambda}_k=\frac{d}{dt}\left(\frac{d\lambda_k}{d\ten{A}}\right)=-\frac{\dot\lambda_k}{\lambda^2_k}\ten{N}_k\otimes\ten{N}_{k}+\sum_{p\ne k}\frac{1}{\lambda_k}\Omega_{kp}(\ten{N}_p\otimes\ten{N}_k+\ten{N}_k\otimes\ten{N}_p)
\end{equation}Then, since $\ten{\dot\lambda}_k=d\ten{\lambda}_k/d\ten{A}:\ten{\dot A}$, comparing with Eq. (\ref{eqAdot}) and taking into account that $\ten{M}_{kk}:\ten{M}_{ij}=\delta_{ik}\delta_{jk}$, which vanishes if $i\ne j$, we have \begin{equation} 
\frac{d\ten{\lambda}_k}{d\ten{A}}\equiv \frac{d^2 {\lambda}_k}{d\ten{A}d\ten{A}}=-\frac{1}{\lambda_k^3}\ten{M}_{kk}\otimes\ten{M}_{kk}+\sum_{p\ne l}\frac{2}{\lambda_k} \frac{1}{\lambda_p^2-\lambda_k^2}\ten{M}_{kp}\overset{s}{\otimes} \ten{M}_{kp}\label{d2l}
\end{equation}Now, we can apply the chain rule as in Eq. (\ref{Lambdak}) \begin{equation}
\frac{d^2\lambda_k^d}{d\ten{A}d\ten{A}}=\sum_{i=1}^{3}\sum_{j=1}^3\frac{\partial^2\lambda_k^d}{\partial\lambda_i\partial\lambda_j}\frac{d\lambda_i}{d\ten{A}}\otimes \frac{d\lambda_j}{d\ten{A}}+\sum_{i=1}^3\frac{\partial\lambda_k^d}{\partial\lambda_i}\frac{d^2\lambda_i}{d\ten{A}d\ten{A}}
\end{equation}
In this expression, $d\lambda_i/d\ten{A}$ is given in Eq. (\ref{dlambda}), $\partial\lambda_k^d/\partial\lambda_i$ is given in Eq. (\ref{dlkd}) and  ${d^2\lambda_i}/({d\ten{A}d\ten{A}})$ in Eq. (\ref{d2l}), so only ${\partial^2\lambda_k^d}/({\partial\lambda_i\partial\lambda_j})$ need to be computed, which is immediately done from Eq. (\ref{dlkd}) 

\begin{align} 
\frac{\partial^2\lambda_k^d}{\partial\lambda_i\partial\lambda_j}&=J^{-\frac{1}{3}}\left(-\frac{1}{3}\frac{\delta_{kj}}{\lambda_i}
+\frac{2}{3}\frac{\lambda_k}{\lambda_i^2}\delta_{ij}\right)-\frac{J^{-\frac{1}{3}}}{3\lambda_j}\left( \delta_{ik}-\frac{1}{3}\frac{\lambda_k}{\lambda_i}\right)\\
&=\frac{J^{-\frac{1}{3}}}{3}\left[2\frac{\lambda_k}{\lambda_i^2}\delta_{ij}-\frac{\delta_{kj}}{\lambda_i}-\frac{\delta_{ki}}{\lambda_j}+\frac{1}{3}\frac{\lambda_k}{\lambda_i\lambda_j} \right]
\end{align} 

An alternative to obtain the tangent is via the fictitious tensor $\ten{\bar S}=2d\Psi/d\ten{C}^d$ where the eigenvalues are $\bar S_i=(1/\lambda_i^{d})d\Psi /d\lambda_i^d  $ --- also written in terms of the manifold--- and $\ten{C}^d:=J^{-\frac{2}{3}}\ten{C}$ is the isochoric right Cauchy-Green deformation tensor. Then
\begin{equation} 
\frac{d\ten{C}^d}{d\ten{C}}=:J^{-\frac{2}{3}}\mathbb{\bar P}=J^{-\frac{2}{3}}\left(\mathbb{I}^S-\tfrac{1}{3}\ten{C}\otimes\ten{C}^{-1}\right)=J^{-\frac{2}{3}}
\mathbb{I}^S-\tfrac{1}{3}\ten{C}^d\otimes\ten{C}^{-1}
\end{equation} 
so
\begin{equation} 
\ten{S}=pJ\ten{C}^{-1}+2\frac{d\Psi}{d\ten{C}^d}:\frac{d\ten{C}^d}{d\ten{C}}
=\left( pJ-\frac{2}{3}\frac{d\Psi}{d\ten{C}^d}:\ten{C}^d\right) \ten{C}^{-1}+2J^{-\frac{2}{3}}\frac{d\Psi}{d\ten{C}^d}
\end{equation}
Rename $\Lambda_i:=(\lambda_i^d)^2$ as the eigenvalues of $\ten{C}^d$; the eigenvalues of $d\Psi/d\ten{C}^d$ are $\partial\Psi/\partial\Lambda_i$ whereas both have the same eigenvectors. Note that $\Lambda_1\Lambda_2\Lambda_3=1$, so only two $\Lambda_i$ are independent. The third term, for example, may be written as 
\begin{align} \frac{\partial\Psi}{\partial\Lambda_3}
\Lambda_3=\frac{\partial\Psi}{\partial\Lambda_1}\frac{\partial\Lambda_1}{\partial\Lambda_3}\Lambda_3+\frac{\partial\Psi}{\partial\Lambda_2}\frac{\partial\Lambda_2}{\partial\Lambda_3}\Lambda_3=-\frac{\partial\Psi}{\partial\Lambda_1}\frac{\Lambda_3}{\Lambda_3^2\Lambda_2}
-\frac{\partial\Psi}{\partial\Lambda_2}\frac{\Lambda_3}{\Lambda^2_3\Lambda_1}=-\frac{\partial\Psi}{\partial\Lambda_1}\Lambda_1
 -\frac{\partial\Psi}{\partial\Lambda_2}\Lambda_2
 \end{align}  so ---note that this is essentially the trace of a rotated Kirchhoff stress for a isochoric deformation, known to be traceless;\ see also \cite{Bonet}, Sec. 5.5.1 for a different explanation
 \begin{equation}
 \frac{d\Psi}{d\ten{C}^d}:\ten{C}^d
 =\frac{\partial\Psi}{\partial\Lambda_1}\Lambda_1
 +\frac{\partial\Psi}{\partial\Lambda_2}\Lambda_2
 +\frac{\partial\Psi}{\partial\Lambda_3}\Lambda_3
 =0
 \end{equation}
Then, the stress takes a typical form and the tangent is derived as usual, see for example \cite{Bonet}, Example 6.8 in \cite{Holzapfel} or Appendix A of \cite{Miehe94_hyperelasticity}, so further details are omitted.  
\end{appendix}
\section*{References}
\bibliography{wypich}

\begin{thebibliography}{10}
\expandafter\ifx\csname url\endcsname\relax
  \def\url#1{\texttt{#1}}\fi
\expandafter\ifx\csname urlprefix\endcsname\relax\def\urlprefix{URL }\fi
\expandafter\ifx\csname href\endcsname\relax
  \def\href#1#2{#2} \def\path#1{#1}\fi

\bibitem{mark2007rubberlike}
J.~E. Mark, B.~Erman, Rubberlike elasticity: a molecular primer, Cambridge
  University Press, 2007.

\bibitem{benitez2017mechanical}
J.~M. Ben{\'\i}tez, F.~J. Mont{\'a}ns, The mechanical behavior of skin:
  {S}tructures and models for the finite element analysis, Computers \&
  Structures 190 (2017) 75--107.

\bibitem{bathe2006finite}
K.-J. Bathe, Finite element procedures, 2nd Ed., Klaus-J{\"u}rgen Bathe, 2014.

\bibitem{treloar1975physics}
L.~R.~G. Treloar, The physics of rubber elasticity, Oxford University Press,
  USA, 1975.

\bibitem{bergstrom2015mechanics}
J.~S. Bergström, Mechanics of solid polymers: theory and computational
  modeling, William Andrew, 2015.

\bibitem{latorre2017understanding}
M.~Latorre, E.~De~Rosa, F.~J. Mont{\'a}ns, Understanding the need of the
  compression branch to characterize hyperelastic materials, International
  Journal of Non-Linear Mechanics 89 (2017) 14--24.

\bibitem{ogden2004fitting}
R.~Ogden, G.~Saccomandi, I.~Sgura, Fitting hyperelastic models to experimental
  data, Computational Mechanics 34~(6) (2004) 484--502.

\bibitem{ibanez2017data}
R.~Iba{\~n}ez, D.~Borzacchiello, J.~V. Aguado, E.~Abisset-Chavanne, E.~Cueto,
  P.~Ladev{\`e}ze, F.~Chinesta, Data-driven non-linear elasticity: constitutive
  manifold construction and problem discretization, Computational Mechanics
  60~(5) (2017) 813--826.

\bibitem{ibanez2018hybrid}
R.~Ib{\'a}{\~n}ez, E.~Abisset-Chavanne, D.~Gonz{\'a}lez, J.-L. Duval, E.~Cueto,
  F.~Chinesta, Hybrid constitutive modeling: data-driven learning of
  corrections to plasticity models, International Journal of Material Forming
  (2018) 1--9.

\bibitem{ibanez2018manifold}
R.~Ibanez, E.~Abisset-Chavanne, J.~V. Aguado, D.~Gonzalez, E.~Cueto,
  F.~Chinesta, A manifold learning approach to data-driven computational
  elasticity and inelasticity, Archives of Computational Methods in Engineering
  25~(1) (2018) 47--57.

\bibitem{sussman2009model}
T.~Sussman, K.-J. Bathe, A model of incompressible isotropic hyperelastic
  material behavior using spline interpolations of tension--compression test
  data, Communications in Numerical Methods in Engineering 25~(1) (2009)
  53--63.

\bibitem{latorre2013extension}
M.~Latorre, F.~J. Mont{\'a}ns, Extension of the {S}ussman--{B}athe spline-based
  hyperelastic model to incompressible transversely isotropic materials,
  Computers \& Structures 122 (2013) 13--26.

\bibitem{latorre2014you}
M.~Latorre, F.~J. Mont{\'a}ns, What-{Y}ou-{P}rescribe-is-{W}hat-{Y}ou-{G}et
  orthotropic hyperelasticity, Computational Mechanics 53~(6) (2014)
  1279--1298.

\bibitem{crespo2017wypiwyg}
J.~Crespo, M.~Latorre, F.~J. Mont{\'a}ns, {WYP}i{WYG} hyperelasticity for
  isotropic, compressible materials, Computational Mechanics 59~(1) (2017)
  73--92.

\bibitem{crespo2018continuum}
J.~Crespo, F.~J. Mont{\'a}ns, A continuum approach for the large strain finite
  element analysis of auxetic materials, International Journal of Mechanical
  Sciences 135 (2018) 441--457.

\bibitem{latorre2018continuum}
M.~Latorre, M.~Mohammadkhah, C.~K. Simms, M.~F. J, A continuum model for
  tension-compression asymmetry in skeletal muscle, Journal of the mechanical
  behavior of biomedical materials 77 (2018) 455--460.

\bibitem{de2017capturing}
E.~De~Rosa, M.~Latorre, F.~J. Mont{\'a}ns, Capturing anisotropic constitutive
  models with {WYPiWYG} hyperelasticity; and on consistency with the
  infinitesimal theory at all deformation levels, International Journal of
  Non-Linear Mechanics 96 (2017) 75--92.

\bibitem{latorre2018experimental}
M.~Latorre, F.~J. Mont{\'a}ns, Experimental data reduction for hyperelasticity,
  Computers \& Structures (in press, doi: 10.1016/j.compstruc.2018.02.011).

\bibitem{arruda1993three}
E.~M. Arruda, M.~C. Boyce, A three-dimensional constitutive model for the large
  stretch behavior of rubber elastic materials, Journal of the Mechanics and
  Physics of Solids 41~(2) (1993) 389--412.

\bibitem{treloar1944stress}
L.~Treloar, Stress-strain data for vulcanized rubber under various types of
  deformation, Rubber Chemistry and Technology 17~(4) (1944) 813--825.

\bibitem{miehe2004micro}
C.~Miehe, S.~G{\"o}ktepe, F.~Lulei, A micro-macro approach to rubber-like
  materials—part {I}: the non-affine micro-sphere model of rubber elasticity,
  Journal of the Mechanics and Physics of Solids 52~(11) (2004) 2617--2660.

\bibitem{kaliske1999extended}
M.~Kaliske, G.~Heinrich, An extended tube-model for rubber elasticity:
  statistical-mechanical theory and finite element implementation, Rubber
  Chemistry and Technology 72~(4) (1999) 602--632.

\bibitem{khiem2016analytical}
V.~N. Khi{\^e}m, M.~Itskov, Analytical network-averaging of the tube model:
  Rubber elasticity, Journal of the Mechanics and Physics of Solids 95 (2016)
  254--269.

\bibitem{urayama2006experimentalist}
K.~Urayama, An experimentalist's view of the physics of rubber elasticity,
  Journal of Polymer Science Part B: Polymer Physics 44~(24) (2006) 3440--3444.

\bibitem{marckmann2006comparison}
G.~Marckmann, E.~Verron, Comparison of hyperelastic models for rubber-like
  materials, Rubber chemistry and technology 79~(5) (2006) 835--858.

\bibitem{shariff2000strain}
M.~Shariff, Strain energy function for filled and unfilled rubberlike material,
  Rubber chemistry and technology 73~(1) (2000) 1--18.

\bibitem{ogden1997non}
R.~W. Ogden, Non-linear elastic deformations, Courier Corporation (Dover),
  1997.

\bibitem{kawabata1981experimental}
S.~Kawabata, M.~Matsuda, K.~Tei, H.~Kawai, Experimental survey of the strain
  energy density function of isoprene rubber vulcanizate, Macromolecules 14~(1)
  (1981) 154--162.

\bibitem{Kawamura}
T.~Kawamura, K.~Urayama, S.~Kohjiya, Multiaxial deformations of end-linked
  poly(dimethylsiloxane) networks. 1. phenomenological approach to strain
  energy density function, Macromolecules 34 (2001) 8252--8260.

\bibitem{amores2019ADES}
V.~J. Amores, J.~M. Ben\'{i}tez, F.~J. Mont\'{a}ns, Average-chain behavior of
  isotropic incompressible polymers obtained from macroscopic experimental
  data. {A} simple structure-based {WYP}i{WYG} model in {J}ulia language,
  Advances in Engineering Software 130 (2019) 41--57.

\bibitem{bavzant1986efficient}
P.~Ba{\v{z}}ant, B.~Oh, Efficient numerical integration on the surface of a
  sphere, ZAMM-Journal of Applied Mathematics and Mechanics/Zeitschrift f{\"u}r
  Angewandte Mathematik und Mechanik 66~(1) (1986) 37--49.

\bibitem{weinert2013fast}
H.~L. Weinert, Fast compact algorithms and software for spline smoothing,
  Springer, 2013.

\bibitem{eilers1996flexible}
P.~H. Eilers, B.~D. Marx, Flexible smoothing with b-splines and penalties,
  Statistical Science (1996) 89--102.

\bibitem{o1986automatic}
F.~O'Sullivan, B.~S. Yandell, W.~J. Raynor~Jr, Automatic smoothing of
  regression functions in generalized linear models, Journal of the American
  Statistical Association 81~(393) (1986) 96--103.

\bibitem{eubank1999nonparametric}
R.~L. Eubank, Nonparametric regression and spline smoothing, CRC press, 1999.

\bibitem{Lewis_silicone}
F.~M. Lewis, The science and technology of silicone rubber, Rubber Chemistry
  and Technology 35~(5) (1962) 1222--1275.

\bibitem{ammar2016effect}
A.~Ammar, Effect of the inverse {L}angevin approximation on the solution of the
  fokker--planck equation of non-linear dilute polymer, Journal of
  Non-Newtonian Fluid Mechanics 231 (2016) 1--5.

\bibitem{nguessong2014new}
A.~N. Nguessong, T.~Beda, F.~Peyraut, A new based error approach to approximate
  the inverse {L}angevin function, Rheologica Acta 53~(8) (2014) 585--591.

\bibitem{kroeger2015simple}
M.~Kr\"{o}ger, Simple, admissible, and accurate approximants of the inverse
  {L}angevin and {B}rillouin functions, relevant for strong polymer
  deformations and flows, Journal of Non-Newtonian Fluid Mechanics 223 (2015)
  77--87.

\bibitem{itskov2012taylor}
M.~Itskov, R.~Dargazany, K.~H{\"o}rnes, Taylor expansion of the inverse
  function with application to the {L}angevin function, Mathematics and
  Mechanics of Solids 17~(7) (2012) 693--701.

\bibitem{jedynak2017new}
R.~Jedynak, New facts concerning the approximation of the inverse {L}angevin
  function, Journal of Non-Newtonian Fluid Mechanics 249 (2017) 8--25.

\bibitem{marchi2015error}
B.~C. Marchi, E.~M. Arruda, An error-minimizing approach to inverse {L}angevin
  approximations, Rheologica Acta 54~(11-12) (2015) 887--902.

\bibitem{darabi2015simple}
E.~Darabi, M.~Itskov, A simple and accurate approximation of the inverse
  {L}angevin function, Rheologica Acta 54~(5) (2015) 455--459.

\bibitem{benitez2018simple}
J.~M. Benitez, F.~J. Mont{\'a}ns, A simple and efficient numerical procedure to
  compute the inverse langevin function with high accuracy, Journal of
  Non--Newtonian Fluids Mechanics 261 (2018) 153--163.

\bibitem{riande1999polymer}
E.~Riande, R.~Diaz-Calleja, M.~Prolongo, R.~Masegosa, C.~Salom, Polymer
  viscoelasticity: stress and strain in practice, CRC Press, 1999.

\bibitem{james1943theory}
H.~M. James, E.~Guth, Theory of the elastic properties of rubber, The Journal
  of Chemical Physics 11~(10) (1943) 455--481.

\bibitem{anthony1942equations}
R.~L. Anthony, R.~H. Caston, E.~Guth, Equations of state for natural and
  synthetic rubber-like materials. i. unaccelerated natural soft rubber, The
  Journal of Physical Chemistry 46~(8) (1942) 826--840.

\bibitem{williams1971polymer}
D.~J. Williams, Polymer science and engineering, Prentice-Hall international
  series in the physical and chemical engineering sciences, 1971.

\bibitem{Bonet}
J.~Bonet, R.~Wood, Nonlinear continuum mechanics for finite element analysis,
  Cambridge University Press, Cambridge, 1997.

\bibitem{Holzapfel}
G.~Holzapfel, Nonlinear Solid Mechanics. A Continuum Approach for Engineering,
  Wiley, Chichester, 2000.

\bibitem{Miehe94_hyperelasticity}
C.~Miehe, Aspects of the formulation and finite element implementation of large
  strain isotropic elasticity, International Journal for Numerical Methods in
  Engineering 37~(12) (1994) 1981--2004.

\end{thebibliography}

\end{document}